# REVIEW
**RESEARCH MEMOIR**

## A review of 70 years with astrometry
**From meridian circles to Gaia and beyond**

By Erik Høg, Niels Bohr Institute, Copenhagen University, Copenhagen, Denmark
Email: ehoeg@hotmal.dk

**Abstract:** In 1953 I heard of an experiment in 1925 by Bengt Strömgren where he observed transit times with the meridian circle at the Copenhagen University Observatory measuring the current in a photocell behind slits when a star was crossing. In 1954 just 22 years old I was given the task as a student to make first test observations with a new meridian circle of the observatory. I became fascinated by the instrument and by the importance of astrometry for astronomy. Work at four meridian circles, two in Denmark, one in Hamburg, one in Lund, and Pierre Lacroute's vision of space astrometry in France had by 1973 created the foundation for development of the Hipparcos satellite, and Gaia followed. In 2013 I proposed a successor satellite which has gained momentum especially thanks to the efforts of David Hobbs and it has a good chance to be launched by ESA about 2045. - But 70 years ago, optical astrometry was considered a dying branch of astronomy, unattractive compared with astrophysics. The following growth built on the still active interest in astrometry in Europe in those years and it was supported by ESA, the European Space Agency. – This review is only about astrometry where I was personally involved.

### 1. Introduction

Here follows a review of 70 years with *astrometry*, which is the branch of astronomy where positions of stars are measured very accurately. From positions obtained over years the motions of stars and their distances are derived. The other branch of astronomy is *astrophysics* where the physical nature of stars and the universe are studied especially by observations of spectra and luminosities of stars and using results from astrometry. The term astrometry has come into use to distinguish from astrophysics since about 1900 when the physical laws of heat and atoms, i.e., thermodynamics and quantum mechanics, were established.

Astrometry has been pursued since Antiquity, and Tycho Brahe (1546-1601) started the modern development by his observations during twenty years from the then Danish Island Hven between Denmark and Sweden. He was a great *astrometrist* when he measured the positions of 1000 stars, the planets, several comets, and the new star in 1572, and he was an *astrophysicist* when he concluded from the observed positions that Stella Nova was further away than the Moon, that it was a fixed star in the eighth celestial sphere where it was believed since ancient time that no changes could occur. Tycho therefore called it "the greatest wonder since the creation of the world".

## 2. Overview

The review is mainly of astrometry where I have been personally involved. In the same period of 70 years, other areas of astrometry were developed and had great significance. The other areas were photographic and CCD astrometry from the ground, radio astrometry, and astrometry from the Hubble Space Telescope. The entire history of astrometry has been presented by Perryman (2012) in a very readable overview of all areas of astrometry, including references.

A constant worry when thinking and writing about all those years which I still remember so well, is that I can mention but a few of all those who meant so much for the projects and for myself personally, when this paper shall be readable and comprehensible. Three reports in Høg (2018a) treat various aspects of the period 1964-1980 more deeply.

I divide the period in two parts: During the 20 years 1953-1973 a strong basis for space astrometry was created thanks to meridian circles in Copenhagen, Brorfelde, Hamburg, and Lund, and to the great vision of Pierre Lacroute in France about astrometry from space. During the second part of 50 years 1974-2023, space astrometry was developed, and two European satellites, Hipparcos and Gaia, were launched by the European Space Agency. This was building on the astrometric tradition in Europe and on the support from ESA. In 2013 a successor for Gaia was proposed and has since been developed so that a launch in 2045 seems probable.

## 3. Copenhagen and Brorfelde observatories

My interest in astronomy began about 1946 when I was 13 years old, living in the countryside on the southern Danish Island Lolland, Høg (2017a). The starry sky was overwhelming, not disturbed by any streetlights. About science and telescopes I read with 11, my own telescopes were built by two blacksmiths in the nearby village Frejlev and at a fine mechanical workshop in Nykøbing with me standing by. I ground and polished the mirrors at home, tested with Foucault, coated with silver, observed variable stars, and one night I caught SS Cygni in eruption. Teachers at the school in Nykøbing and astronomers in Copenhagen supported my interests. In 1950 I moved to Copenhagen to study mathematics, physics, chemistry, and astronomy at the University. My initial goal was to become a high-school teacher because I had understood it would be unrealistic to think of becoming a professional astronomer. I became a frequent guest at the Copenhagen Observatory at Øster Voldgade 3 where my dedicated study of astronomy began in 1953.

My 70 years with astrometry began at the observatory in 1953. I heard of an experiment made by Bengt Strömgren (1908-1987) in 1925 with *photoelectric astrometry*, a brilliant idea of a new method for astrometry which has borne fruit, most recently with the Gaia satellite. In the Copenhagen meridian circle, he placed slits and a photocell behind, see figure 1. A quite advanced piece of electronics for that time amplified the photo current so that he could record transit times when a bright star crossed the slits.

In 1933 he described a further idea for obtaining the median transit time, which was crucial for my later ideas, see figure 2. Instead of instantaneous measurements of the photo current from a single photocell, integration of the light from two photocells is introduced resulting in great reduction of the noise. A mirror behind the slits is quickly switched at a known times so that the light is sent to one or the other cell, and each of the cells integrate the light. On five pages an arrangement of apparatus and accurate timing is described in the paper by which the mirror switches when the star is near the middle of each slit.

In this manner the two cells will integrate nearly equal amounts of light, and from these amounts the transit time for the middle of the slit can be computed, which means that the median transit time has been found. At first, I saw this proposal as quite unrealistic, but in fact, this was my inspiration when in 1960 I invented the new method to do astrometry with photon counting where the median time is derived from digitally recorded photon counts.

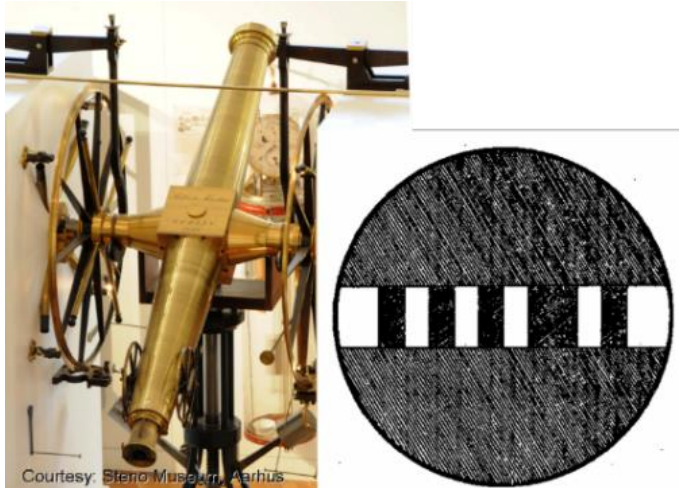

*Figure 1. The meridian circle in the new Copenhagen observatory from 1861 which is now exhibited in the Steno Museum, Aarhus. At right the system of slits used by Bengt Strömgren in 1925 for the first experiments in the world with photoelectric astrometry, see more explanation in section 3. © Steno Museum*

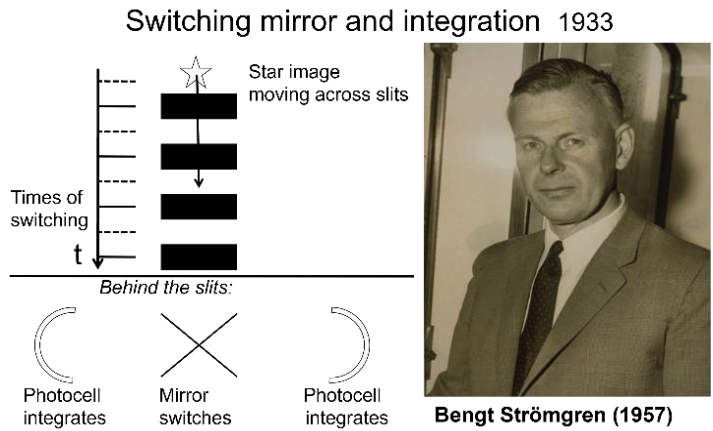

*Figure 2. At left, my explanation of Bengt Strömgren's idea from 1933 which was basis for my invention of photon counting astrometry in 1960, see more in sections 3 and 4. - Figure from the private collection of the author.*

My tutor Peter Naur (1928-2016) took the books off the shelf in the library and showed me the papers, Strömgren (1925, 1926, 1933). In fact, I never discussed it with Strömgren in those years, for him a photographic method was the method to be developed at that time.

Bengt Strömgren became professor of astronomy in 1940. He was one of the great astrophysicists of the century, but he was also a supporter of astrometry. In 1940 he ordered from England a new meridian circle to be the main instrument of the new observatory he intended to build far outside Copenhagen. The instrument was erected in 1953 on a hilltop at the small village Brorfelde 50 km to the west of Copenhagen, see figure 3.

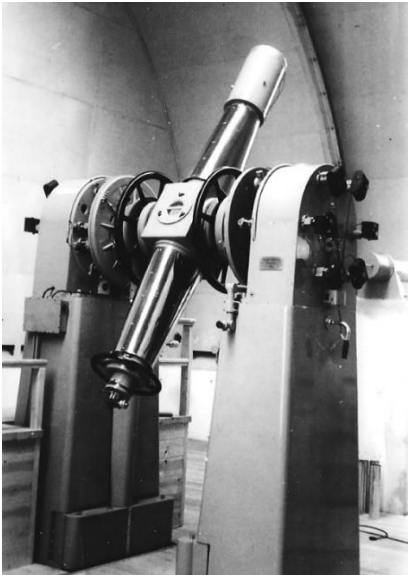

*Figure 3. The Brorfelde meridian circle in 1954 where I became fascinated by astrometry and found the inspiration to my further work on astrometry. - Figure from the private collection of the author.*

As a student of 22 years in 1954 I was sent to work with this instrument. I was quite alone there and sometimes slept in a haystack when clouds came, see figure 4. Much later, I saw in a correspondence (Høg 2015) between Bengt Strömgren and Julie Vinter Hansen (1890-1960) that Peter Naur had said that sending me there was perhaps a sure way to kill the interest in astronomy in a young man – but I was different.

My task was to check the stability of the instrument by taking short exposures on a fixed photographic plate of Polarissima, a star so close to the celestial pole that it remained within the field of the small plate during the whole night. The positions of the star were then measured on a machine in Copenhagen. A very simple study, but I became fascinated by the instrument and by astrometry as the important tool for astronomy.

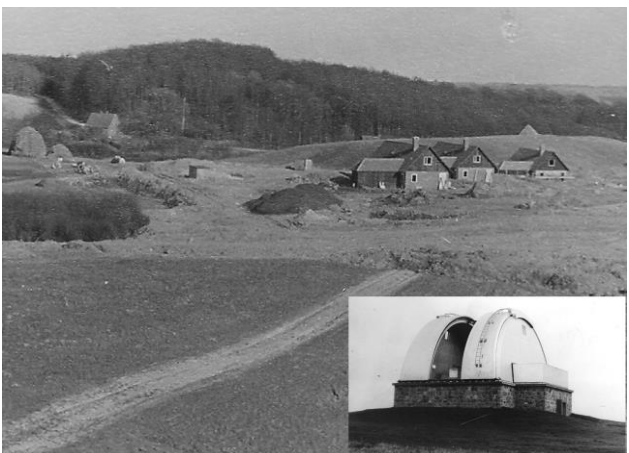

*Figure 4. View from top of the Brorfelde meridian building where I observed quite alone from 1954 and sometimes slept in a haystack when clouds came up. The photo from 1955 shows the empty landscape towards observers' houses, still uninhabited. Insert, the building itself. - Figure from the private collection of the author.*

I attended lectures on astrometry in 1953 by Professor Kaj Aage Strand (1907-2000) at the observatory where he visited from Dearborn Observatory, Evanston, USA. He was born and educated in Denmark and became assistant in Leiden to another Danish astronomer, Professor Einar Hertzsprung (1873-1967), director of Leiden Observatory 1935-44, and famous for his epochal work in 1906 on the best-known diagram in astronomy, the Hertzsprung-Russell diagram (HR-diagram), a relation between absolute brightness and spectral type of stars.

Hertzsprung also supported my interest in astrometry. After retiring from Leiden in 1944, he lived with his daughter in the village Tølløse at the train station closest to Brorfelde and I often visited him there on invitation. We talked about astronomy and astronomers, and he would point at the photo of the optical workshop in Hamburg with Berhard Schmidt (1879-1935) whom he admired. I measured a photographic plate with double star exposures while Hertzsprung noted the numbers, and the visit always ended in the neighboring room where his daughter served us a dessert. He paid my ticket and accompanied me back to the station. Later in Hamburg, I measured double stars with the great 60 cm refractor on his proposal. In 1961 he urged me to attend the IAU General Assembly in Berkeley, California, where he was going himself and he found the funding for my first visit to America where I met great wonders of astronomy and nature.

Much later in Hamburg, I built a photoelectric scanning micrometer (Høg 1971a) for double stars with a small online computer PDP-8/S which was also used to measure the first timing in Europe of the optical Crab pulsar, Høg & Lohsen (1970). Professor Christian de Vegt (1936-2002) measured lunar occultations with this new type of digitized instrument.

The first time I met astrometry at an international meeting was in 1958 at the IAU GA in Moscow. On return to Denmark, I visited the large and famous Pulkovo Observatory near Leningrad where Nikolay Pavlov (1902-1985) showed me his photoelectric micrometer at a transit instrument.

Still in Brorfelde 1957, Peter Naur urged me to go abroad, and I obtained a grant for a ten-month study at the Hamburg Observatory from October 1958, but in fact I stayed 15 years. I wanted of course to become an astrophysicist because I was aware that astrometry was old fashioned. That should be possible in Hamburg-Bergedorf, at that time the largest observatory in Germany with about 50 employees.

### 4. Hamburg Observatory

At first, I worked on an astrophysical project in Hamburg, but in 1960 I had the ideas (Høg 1960) for a new method of astrometry, see figure 5. Slits are placed in the focal plane of the meridian circle and photons are detected by a photo-multiplier tube behind the slits. The photons are counted in short intervals and the counts are recorded on punched tape. These counts are later analyzed in a digital computer and the median transit time for each slit is obtained.

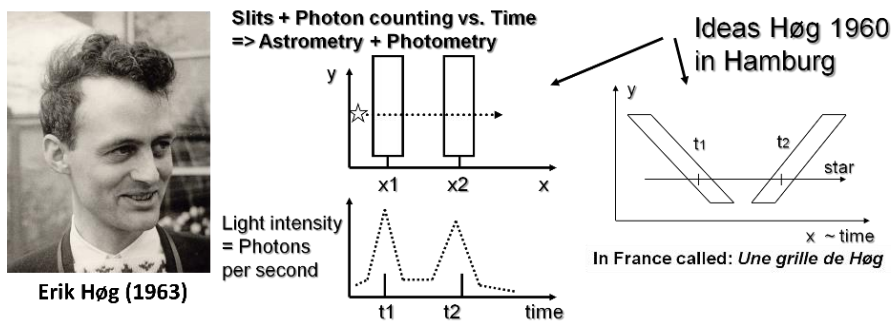

*Figure 5. Photon counting astrometry used slits and a photo-multipler tube behind the slits. The photon counts were recorded on punched tapes which were later treated in a digital computer. This digital technique invented in 1960 was used in the Hipparcos astrometry satellite launched by ESA in 1989. - Figure from the private collection of the author.*

The ideas were described in my first scientific paper (Høg 1960). The 2nd paper from 1961 was also astrometric on the determination of division corrections with a new method soon used at all meridian circles, and the 3rd paper contained measures of visual double stars with the 60 cm refractor.

The proposal was very well received when I mentioned it in the Hamburg Observatory in July 1960, Høg (2017b). The idea came at the right moment of time in Hamburg because the observatory was planning an expedition with the meridian circle to Perth in Western Australia as part of an international collaboration for astrometric observation of the southern sky. The expedition was planned to go with visual observations of the stars, but the director Professor Otto Heckmann (1901-1983) and the expedition leader Dr. Johan von der Heide (1902-1995) decided to use the new photo electric method. It was a risky gamble with many problems, but it was successful in the end.

The method was implemented on the Hamburg meridian circle, Høg (2014c) in the years 1960-67, see figure 6. Some of the electronics I built myself as I had learnt from Peter Naur in Brorfelde, but most of it was built by the AEG company in Berlin where I visited several times to agree on the design.

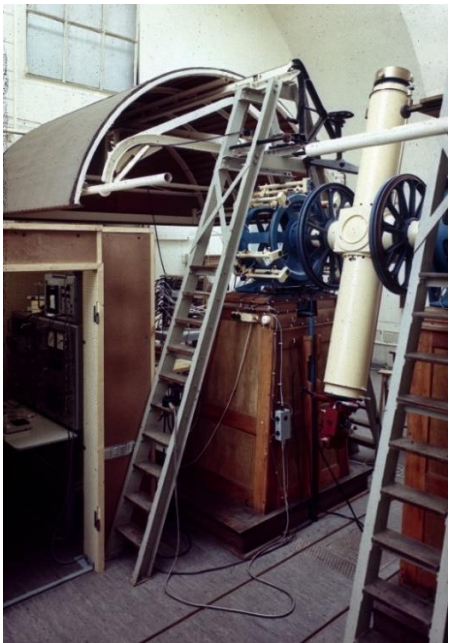

*Figure 6. The Hamburg meridian circle was semi-automatic, and it was the first fully digitized astronomical telescope. The observer standing at the telescope turned the instrument to the declination told him by the assistant sitting in the cabin at left. She, usually his wife, had the star list and she would set some numbers on the electronics which was placed in the heated cabin, protected from the cold air outside. With this instrument 110,000 observations of stars and planets of unprecedented accuracy were obtained during 5 years at the Perth Observatory in Western Australia beginning in 1967. - Figure from the archive of the Hamburg Observatory.*

It was operated for 5 years at the Perth Observatory resulting in a catalogue of 25,000 stars (Høg, von der Heide, et al. 1976 and Høg 1976) with an accuracy of ±150 mas. Up to twelve people took the observations on punched tapes and later reduced the data in the GIER computer, which had been manufactured in Denmark.

The GIER computer was one of the first transistorized computers in the world and ten times faster than the cheapest American IBM computer we could afford, Høg (2017b), any laptop today is several million times faster than GIER was and keeps immediate access to millions of times more data. This speed was crucial for the project since it enabled the observers to keep up with the data reductions as the observations were obtained in the many clear nights at Perth. Estimation of the transit time for each slit was done with the median method, a simple and fast method. Only much later did I learn about the more accurate location estimation by maximum likelihood, but this is more computer expensive and could not have been done by GIER in an acceptable time – perhaps a luck that I was ignorant on that point.

I wrote all the programs very conveniently in ALGOL 60, an algorithmic language just developed at the computer company, Regnecentralen (2023), building the GIER. This development was led by Peter Naur who had left astronomy entirely in 1959 although he had worked with astronomy since he was a boy. Naur and I were used to logarithms and mechanical calculators throughout our time as students, but he pioneered electronics and electronic computers, and I have absorbed this spirit while sharing his office as a student and as his collaborator in Brorfelde until October 1958. He soon became the first professor of computer sciences in Denmark and founded the institute for Datalogy at the Copenhagen University. In 2005 he was awarded the Alan Turing prize, called the Nobel prize in computer sciences.

### 5. Strasbourg Observatory

In France, the new method of photon counting astrometry was adopted as basis for the great vision of a space-based astrometric mission by Professor Pierre Lacroute (1906-1993), director of the Strasbourg Observatory, see figure 7. His work 1964-74 led him to propose a scanning satellite with a split mirror. It was a purely French project to begin with. Everybody could see the advantage of astrometry from outside the atmosphere, but nobody outside France worked on such ideas.

Lacroute presented his first ideas of space astrometry, Lacroute (1967, 1974), at a meeting in Bordeaux in July 1965, according to Kovalevsky (2009). This was the first time that such type of astronomy was proposed for a space mission. The potential advantages were clear: no atmosphere and no gravity, and perhaps thermal stability if that would be technically feasible.

I attended the presentation by Lacroute at the General Assembly of the IAU in Prague in 1967, but to me and others the technical problems seemed utterly underestimated. Also, because he said it could be done for 10 million French francs. The proposal did not start any activity outside France, but people were interested. The interest was fortunately shared by his student Pierre Bacchus (1923-2007) and other French astronomers. It was really Professor Pierre Lacroute's great vision of astrometry from space.

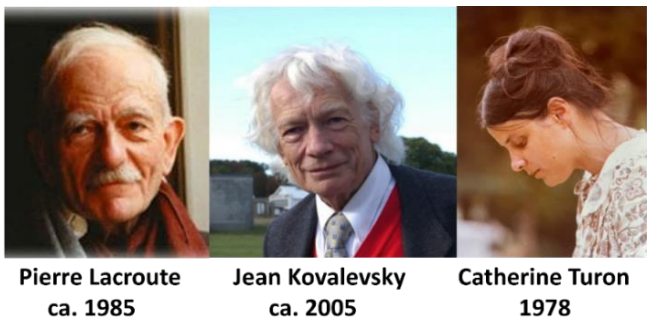

**Pierre Lacroute**          **Jean Kovalevsky**          **Catherine Turon**
**ca. 1985**                 **ca. 2005**                 **1978**

*Figure 7. Three French astronomers with leading roles in Hipparcos. Pierre Lacroute is the initiator of space astrometry, and I believe that his brilliant idea of measuring near to a very fixed angle (basic angle) realized by a beam combiner that reflects the light from two directions into a telescope may very well derive from the impersonal astrolabe of André-Louis Danjon (1890-1967). This instrument incorporates a basic angle for the definition of the instrumentally defined altitude. This became a key for the European success of Space Astrometry. Three more key ideas in Hipparcos and Gaia about the beam combiner and the sky scanning were introduced by me in December 1975 and the following month and they are mentioned at the end of section 14.- Photos from the web and from Catherine Turon.*

In 1970, CNES decided not to pursue any purely French space missions, but only European programs and Pierre Lacroute proposed his projects of space astrometry to ESRO in November 1973. Professor Jean Kovalevsky (1929-2018) was director of a new astrometric observatory near Nice in Southern France. He played a major role by presenting and pushing the project in the ESRO organization (which became ESA in 1975), and then by bringing fundamental contributions to it.

Once presented to ESA, the question was: would such a mission interest a sufficiently important scientific community? That was the question asked, and for this reason Kovalevsky convened a Symposium, ESRO (1975), on Space Astrometry in Frascati near Rome. It was held in October 1974 and gathered 41 participants from Europe and the United States.

In the two days Lacroute presented his ideas. There were many other presentations, one of them by me. At an IAU Symposium in 1973 I had reviewed the modern developments of the meridian circle, Høg (1974a). Now at Frascati, I predicted (Høg 1974b) that automatic meridian circles could compete with the space astrometry predicted by Lacroute. I also said that such instruments could have a bright future. This became true when the automatic meridian circle from Brorfelde was moved to La Palma and observations began in 1984, see section 8 and figure 10.

Catherine Turon Lacarrieu was a young French astrophysicist invited by Jean Kovalevsky and she came to play a leading role in Hipparcos collecting the input catalogue of 120,000 stars to be observed with Hipparcos. In Frascati she presented a first review of the astrophysical consequences of the high improvement in the measurement of trigonometric parallaxes expected from the two versions of space astrometry presented by Pierre Lacroute again in Frascati. The underlined consequences were ranging from luminosity calibration of many more types of stars and the consequences on the cosmic distance scale to age criteria and galactic structure.

A year before the Frascati meeting, something of great significance for the coming space astrometry had happened in the Lund Observatory in Sweden, not far from Copenhagen.

## 6. Lund Observatory

After completion of the Australian expedition, there was no basis for meridian astrometry in Hamburg anymore. But I was needed in Brorfelde where they wanted to introduce photon counting astrometry. I obtained a position as associate professor at the Copenhagen University and left Hamburg in August 1973 with my family of five, all three children were born there.

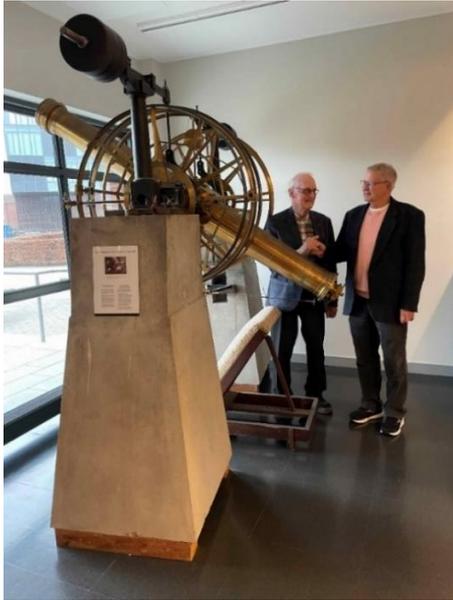

*Figure 8 The Lund Observatory meridian circle now on permanent exhibition, with Lennart Lindegren and me shaking hands in 2023. Astrometry and this instrument had fascinated Lennart in 1973, just as had happened for me at the Brorfelde meridian circle 20 years before when I was as young as he was. - Figure from the private collection of Lennart Lindegren.*

Very soon I heard that a young student in Lund Observatory was working at the old meridian circle there. I wondered because what could one person do when I knew that a whole staff was needed to operate a meridian circle? A meeting was arranged, and I met Lennart Lindegren who was then 23 years. He showed me the Merdian Circle (figure 8) and some impressive improvements he had made. He was aware that he could not make the instrument operative, but he had become fascinated by the fine mechanics of the old instrument and by astrometry, he said. So, he wanted to make his PhD in astrometry, a quite unusual wish in those years and his teachers in Lund wondered, but they fully supported what then happened.

I gave him the observations of five major planets and four minor ones from the Meridian Circle in Australia, and he made a brilliant analysis (Lindegren & Høg 1977) for his thesis. He took physical limb darkening into account, not only the classical geometric darkening. Nobody had done that before, and he used available satellite observations of the planets to get the most correct darkening. With Lennart on board in 1973, the stage was set for the great advances in astrometry we have witnessed since and which shall now be my subject. Here ends the first part of my review with the 20 years period 1953-1973 where a strong basis for space astrometry was created.

## 7. Hipparcos astrometry satellite up to 1980

A detailed account of the years 1975 to 1979 leading to Hipparcos is given in Høg (2018c), only part of this is told hereafter. The Hipparcos mission was approved by ESA in 1980 and this decision changed everything: We could trust that the mission would become a reality, not just remain a possibility among other proposals for scientific missions, and it became reality in the first try, only five years after the first realistic design had been presented. Had this not happened some of my colleagues would have carried on and tried again as is normal. But I would have left the project because I had seen that the resistance from astrophysicists with competing projects was immense as I have explained in Høg (2018d), and because I was deeply involved in the development of the automatic meridian circle in Brorfelde. In 1980 this project went through a difficult period, and I believed it still needed my efforts. But the observatory found ways to continue without me as reported in the following section, and I could concentrate on space astrometry.

Back to October 1974, the conclusion of the meeting in Frascati was written by Jean Kovalevsky and says that further studies should be carried out in 1975, for that purpose a study group should be set up. This happened and in a telephone call from Paris I was invited to join the group.

In fact, I was very skeptical about Lacroute's proposals because his instrument designs seemed quite unrealistic. But of course, I could see the astrometric advantage of observations from outside the Earth's atmosphere. Despite my skepticism, I decided to join the group since it was my duty to do so when ESA asked. The referee says that he had the same initial skepticism but paired with great expectations. Consequently, he took part in ESRO's Space Astrometry Symposium in Frascati in 1974, and he has since been pushing Dutch participation because if it could be done it would be "the breakthrough, we all would need to be part of."

At the first meeting of the Mission Definition Group in October 1975, Lacroute presented the ideas we had heard in Frascati including two projects. The first proposal utilized a 40 cm telescope to be flown by Spacelab in 8 missions spread over 2 years to measure 2000 stars within 0.000,5 arcsec for the five astrometric parameters and 40,000 stars with less accuracy. In those years NASA's Space Shuttle was sold to us as a cheap way to space, but 8 missions for one project was quite extravagant, and the Space Shuttle soon became a very expensive access to space. The second project was a scanning satellite called the TD option with two telescopes on a free flyer like the TD-1 satellite in a slow spin, shown in figure 2 of Høg (2018c). It could measure about 300,000 stars and obtain about 0.001 arcsec for bright stars and <0.003 arcsec for m<10 mag. Both projects were based on a very stable reference angle implemented by a *beam combiner*, a mirror system that sent light from two fields on the sky into the same optical telescope. The detection was done by photon counting photomultipliers behind systems of slit.

I asked the chairman at our meeting which of the two proposals by Lacroute we should study first. The chairman encouraged us to forget the existing proposals and just think how we could best use space technology for our science. These wise words from the chairman, Vittorio Manno (1938-2022), changed my mind completely. I had never been interested in space technology, but I returned to Brorfelde in a new mood. In six weeks, I made a new design in a report of an astrometric satellite where I adopted Lacroute's ideas of a beam combiner in a scanning satellite and introduced 10 new design features or ideas.

I sent my proposal to ESA in December 1975 (section 3.1 in Høg 2018c) calling it "TYCHO". Soon after, I was invited to visit ESTEC for a couple of days. I was tutored by the engineer Maurice Schuyer on, e.g., low earth orbits. I explained the *revolving scanning* required for my design *with the rotation axis at fixed angle to the direction to the Sun*. I remember that they could not understand what I meant, they said that such a scanning motion had never been tried before. But when Hipparcos studies began these engineers soon understood everything very well. The scanning is illustrated in Gaia (2023) and described by Lindegren & Bastian (2011).

The new chairman Mr. R. Pacault said to me in February 1976 that I should better find another name than TYCHO. So, I called it Option A which it kept also during the subsequent Phase A Study. The Greek astronomer Hipparchos had been suggested as name father, and Jean Kovalevsky had introduced the acronym HIPPARCOS, which was later changed to Hipparcos. Nothing was decided during Phase A, but the mission was suddenly named *Hipparcos* at the approval by ESA in 1980. I later asked Jean why the name could not be TYCHO and he answered that a neutral name was better when so many countries were involved.

Some of the 10 new design features are shown in figure 9 (the same as in figure 3 of Høg 2018c). The beam combiner must have *a carefully chosen basic angle, e.g. not 90 degrees* as in Lacroute's proposal. It must have *only two plane surfaces* and *the scanning must be one-dimensional*, not two-dimensional as Lacroute always preferred, even in Option B of March 1976 (see section 3.2 in Høg 2018c). An image-dissector tube (IDT) behind a modulating grid is used for the photon detection instead of photomultipliers behind slits, this improves the observing efficiency by a factor about 100.

This large factor shall be explained: Obviously, a star gives us no information about it position when it moves in the long space between the slits. Introduction of an image-dissector tube as the main detector brings a scanning small sensitive spot, capable of tracking any chosen star to be observed. The grid is modulating the light because a star is touching the edge of a slit all the time and this modulation contains the astrometric information.

An input catalogue of 100,000 carefully chosen stars should be observed so that all observing time by the IDT would be concentrated on those stars. This was made possible by introduction of *star mapper slits*. The slow spin of the satellite was a special "revolving scanning" giving a uniform scanning of the entire sky, this required *active attitude control* instead of the passive control preferred by Lacroute. It is interesting to note that the new design in 1975 was based on a technology for detection, the IDT, which had been available also ten years earlier if somebody would have thought of combining it to an astrometric mission.

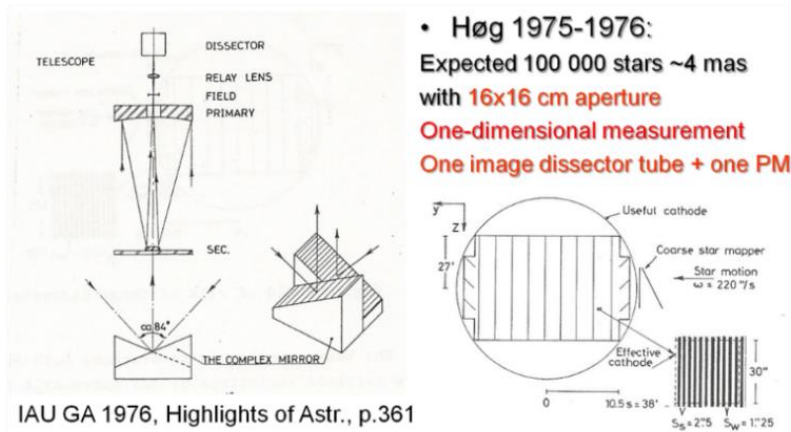

IAU GA 1976, Highlights of Astr., p.361

*Figure 9. Left, my design as of 1976 of the optical system for an astrometric satellite, later called Hipparcos. Light from two fields on the sky are reflected by the mirror (beam combiner) with two plane faces onto a shared focal plane, and the rigid mirror block gave that the two fields a very stable angle to each other, but the one milli-arcsecond accuracy was only obtained thanks to frequent calibration and monitoring as we learned later. The two fields follow each other on the sky as the satellite is slowly rotating and the stars are measured in the focal plane. At right, the expected performance, and the focal plane layout. An image dissector tube is placed behind a modulating grid, and a photomultiplier tube detects the light from stars behind a coarse star mapper. - Figure from the private collection of the author.*

My proposal, soon called Option A, was simpler and more efficient than Lacroute's original ones. Lacroute followed in March 1976 with a new proposal Option B, shown in figure 4 of Høg (2018c) in which some of my ideas were included but much was very different, e.g., it had two-dimensional scanning, it had a beam combiner with three directions of view, it did not use an input catalogue. For details about these options and the Phase A study, see the same reference.

My proposal was well received in the ESA study group. But I could see that the observations would lead to a system with 10 million equations and 500,000 unknowns to be solved by the method of least-squares. This was a formidable task for the available computers at that time. In 1976 Lennart Lindegren was just two years into his PhD studies (completed in 1980) when I posed him the problem. It took him only four weeks to write all the equations in a nicely typed report. It was the three-step method later used on the real data from Hipparcos.

At a meeting soon after, in the ESA study team for Phase A where they had met Lennart, our British colleague Andrew Murray (1926-2012) said to me: The best you have ever done for astronomy was to find Lennart. And I fully agreed as presented in Høg (2008). Without Lennart there would have been no Hipparcos approval in 1980, and probably never.

I was very happy to see what Lennart wrote on our meeting in 1973 in his autobiography for the Shaw Prize he received in 2022: "The Danish astronomer Erik Høg had just returned to his native country after 15 years at the Hamburg Observatory, where he had developed a new technique for meridian observations, photon-counting astrometry. Working at Copenhagen University Observatory, within 100 km of Lund, he was therefore a world-leading expert in precisely the field that I had just chosen for my thesis! Meetings were arranged, and Erik became my de facto PhD supervisor, colleague, and friend. In 1976 he introduced me to the Hipparcos project and the following year I joined him as a member of the space astrometry study team set up by the European Space Agency (ESA)."

The proposal for space astrometry was selected by ESA in July 1976 for a so-called Phase A study. Teams for the studies were selected at a meeting of the Astronomy Working Group on 8/9 Dec. 1976 in Paris, there were 12 members in the team for astrometry, 7 astronomers and 5 from ESA, see section 4 of Høg (2018c) for more detail. It was not at all easy for me to get Lennart on the team, at a critical point I said that if only one of us could get into the team it had to be Lennart, see Høg (2008) p.3. - A great paper from those years on photoelectric astrometry: Lindegren (1978).

A key aspect of getting a mission like this adopted was to build support for Hipparcos throughout the scientific community, and that gained high speed after the Phase A was approved. Studies were made and international meetings took place during 1978 and 1979, e.g., in Padua, Italy, June 5-7, 1978: Colloquium on European satellite astrometry. The proceedings by Barbieri & Bernacca (1979) lists 47 participants and 32 contributions on 303 pages.

The use of an input catalogue was decided by the study team in November 1977. This was only needed for my Option A, so back in Denmark I could started my planned "inquiry on projects", see Høg (1979). I then visited several astronomical institutes especially in Germany to give talks about the project. I also distributed to astronomers mostly in ESA countries a questionnaire asking for their favorite scientific projects and how many stars they would like to have observed by the astrometry satellite. Catherine Turon did similarly in France, Switzerland, and Spain. These answers were starting point for the Hipparcos Input Catalogue. This catalogue (Turon et al., ESA 1992) was compiled several years later by the Input Catalogue Consortium based on 214 proposals submitted in answer to the Invitation for Proposals issued by ESA in 1982

The satellite design submitted for approval was essentially Option A, but this was never said explicitly, only the design features were discussed. Hipparcos was approved in 1980 in hard competition with astrophysical projects see Høg (2018d), ESA (2000a) and Høg (2011a, 2011b).

The circumstances leading to Hipparcos and its approval in 1980 are discussed in Høg (2018b), in the form of interviews from 2017 with scientists about how the mission was conceived up to the begin of technical development. It is based on correspondence with astronomer colleagues and a wise historian of science, Professor John L. Heilbron (1934-2023) in Berkeley.

Hipparcos was the first satellite to obtain "high-precision global absolute astrometry from space", and Gaia is the second spacecraft to do so. These satellites provide very strong astrometric foundation for all branches of astrophysics from the solar system to quasars, a foundation which must be kept up to date as astrophysics is rapidly developing, and it can only be done by such satellites. It is therefore interesting, for me even scary, to think of how much the creation of Hipparcos leading to the approval in 1980 depended on a mere handful of astronomers, see figure 5 in Høg (2011a), during the preceding fifty-five years since 1925. This dependence on so few in the beginning is a fact for me who has witnessed and taken active part in the development of astrometry since 1953. Given the state of the field and recruitment and funding mechanisms, it is very unlikely that it could have happened if anyone of these few persons had been missing.

The report contains interviews with a dozen persons on this question. I try to convey the evidence for this dependence as I have done in numerous reports before. By now, "astrometry has been regained" from a state of weakening or slow improvement before Hipparcos, and with Gaia, astrometry has even become an "almost respectable pursuit" as a colleague has said. This must be maintained in the future for the sake of astronomy and astrophysics.

Imagine the decision in 1980 had been negative and remember that for decades up to 1980 the astrometry community was becoming ever weaker, the older generation retired and very few young scientists entered the field. I personally would have lost the faith that the astrophysicists would ever let such a space mission through, and others would also have left the field of space astrometry. If someone then would have tried a revival of the idea one or two decades later, the available astrometric competence would have been weaker, and where should the faith in space astrometry have come from? When Hipparcos became a European project in 1975, however, and the hopes were high for a realization, the competence from many European countries gathered and eventually was able to carry the mission. This could not have been repeated after a rejection of the mission.

The referee kindly shared some insights in the report, which I do not remember, and which deserve to be quoted here. The referee admits that he speaks out of personal memory: "After the completion of the Phase A report, late in 1979 the SSAC advised to stick to the policy that ESA should provide the platforms, and the adhering member institutes the experiments/payloads. We, the Hipparcos Science Team, perceived that as a great threat and thought of making the expenditure of the data treatment and catalogue publication as a sound counterargument. We even had a meeting on the construction of a consortium (the later consortia) in Rome to better elaborate this effort and to attach a budget to it. We arrived at a budget like the expenditure needed for ESA for funding the payload too.

Henk van de Hulst agreed to the strategic weight of this argument and invited me to the SPC meeting. There he gave me the floor to explain this issue, and with his help it was recognized as a sensible reason to override the SSAC advice and fund the payload too. A necessity in the absence of any space institute with the required experience to build the Hipparcos payload. The expenditure from the scientific institutes still would match the SSAC required member state support for the experiments from the established space institutes. It is my conviction that Henk van de Hulst this way made a significant contribution to bringing Hipparcos to life.

At the end of that paragraph, I think the statement that we would have lost this last opportunity if the decision in 1980 would not have been made as it was, does not do full credit to what the phase A team had achieved. It had greatly elaborated the way to do precision measuring, beyond the boundaries of astrometry. And it had attracted very good younger people like, e.g., Lennart Lindegren. I think the phase A effort was the key trigger for a new lease on life for astrometry.

Agreed a different outcome of the 1980 decision would have retarded Hipparcos quite strongly, and Gaia would most likely not yet have been built. And losing the interest of the author would have been detrimental."

Later on the referee added: "… I am afraid I did not do proper justice to the event at the AWG earlier that year (Januari 24th 1980) where under active support of Ed van den Heuvel Hipparcos already made a 8 : 5 win for priority in ESA's Science program. So, without taking away anything from the contribution by Henk van de Hulst the preparatory success of Ed van den Heuvel's contribution merits better mention than I gave."

My own pessimism about this distant past has not changed after having heard the opinion by others including the referee because I have seen the problems throughout those years and I still remember well, but I certainly have great optimism about the future for astrometry based on the new foundation from the two successful space astrometry missions. Let me conclude with a question: Could NASA have realized an Hipparcos-like mission? NO! For two reasons: the American astrometric community had much less expertise to draw from than were available in Europe, and secondly, as an American colleague said to me in those years: "You can convince a US Congressman that it is important to find life on other planets, but not that it is important to measure a hundred thousand stars."

## 8. The Glass Meridian Circle and tube refraction

In those years, meridian circles were still in use at many observatories, e.g., in USA, USSR, France, Brazil, Japan, Argentina, Yugoslavia, and Australia to mention some.

In fact, the meridian circle developed in Brorfelde became a great success, especially thanks to the team leader Leif Helmer and Claus Fabricius. It was moved to La Palma (figure 10) in a collaboration between Copenhagen University, Royal Greenwich Observatory, and Instituto y Observatorio de Marina, San Fernando. Routine observations began in May 1984 and continued until 2013, according to Wikipedia (2023) and C. Fabricius (priv. comm. 2023). The first catalogue was published in 1985 and a composite catalogue of all observations until May 1998 is available at Vizier (2001). In 1998 the original slit micrometer was replaced by a CCD camera, see Evans, Irwin & Helmer (2002). The Carlsberg Meridian Telescope on La Palma was the first fully automatic astronomical telescope. From 1997 it was supervised remotely via the Internet, and all data was transmitted to the three countries. In 2005 Spain took over the Danish instrument.

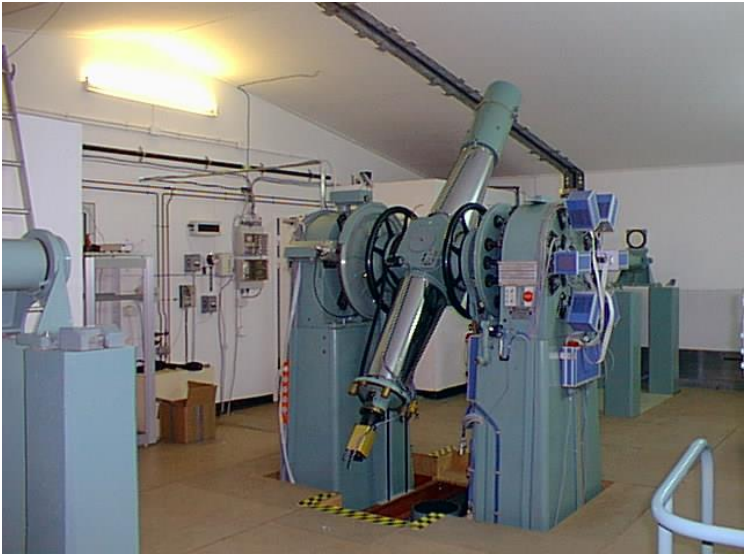

*Figure 10. The Carlsberg Automatic Meridian Circle on La Palma from 1984 after development at Brorfelde. The instrument produced 100,000 observations of stars per year with the slit micrometer as planned and a lot more after 1998 when CCD detectors were introduced. - Figure from the private collection of Dafydd Wyn Evans.*

Development of new types of meridian circles was also going on, e.g., in USSR, in USA, and I pursued a project of this kind. My ideas began in 1967, it should be a horizontal instrument without any long telescope to be tilted to the declination of the star because it had long been suspected that a flexure of the long telescope due to gravity would bend the line of sight relative to the graduated circle, see Høg (1971b, 1973). This "tube flexure" would result in systematic errors of the observed declination, and such errors of declination had been known since about 1900. But no definitive proof had been found that they were due to telescope tube flexure, it was just the general opinion about this "classical" problem of astrometry. During the many years of developing my instrument, I found however quite unexpectedly the true cause of the classical problem of declinations, and I was able to remove the error, as I shall explain later.

I gave many talks at international meetings about my Glass Meridian Circle, GMC, as it was called, see Høg (1971b, 1973) and figure 11.

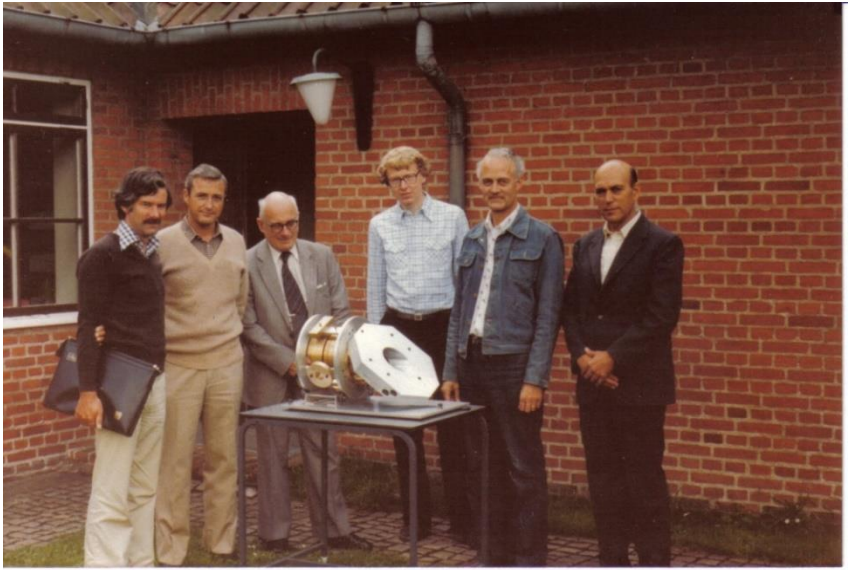

*Figure 11. A model (1:1) of the glass meridian circle at the workshop in Brorfelde in August 1980. Six meridian astronomers from three countries, from left: no 1 and 3: L.V. Morrison and R.H. Tucker from UK, no 2 and 6: J.L. Muiños and L. Quijano from Spain, no 4 and 5: C. Fabricius and E. Høg from Denmark. They were having a meeting in Brorfelde about the common project for the Brorfelde meridian circle on La Palma and used this opportunity to look at the experiments for the future of meridian circles. - Figure from the private collection of Claus Fabricius.*

In 1973 an IAU Study Group for Horizontal Meridian Circles was established and chaired by me until 1991. At the IAU Assembly in Montreal in 1978 a session about horizontal meridian circles was held, and there I met Professor Yeh Shuhua (1927*) "the grand lady of Chinese astronomy".

Mrs. Yeh invited me to a small meeting in Shanghai in September 1980. In the six weeks before I had to be there, I learnt to speak a little Chinese, enough to have great fun with it at all my visits. No agenda was given beforehand, but the travel went through Hongkong and Guangzhou to the old capital Xian in Shaanxi where I was welcomed at the airport by the director of Shaanxi Astronomical Observatory (SAO) and others. We went to the observatory (SAO) in Lintong near Xian, obviously a main destination for my visit. I presented the Hipparcos project, just approved by ESA, and the Glass Meridian Circle as I later did in Beijing and Shanghai. My hosts were very keen to show me the marvels of China, the nature as well as old and new culture, visiting Beijing, Hangzhou and Suzhou, always accompanied by my interpreter.

In Shanghai Mrs. Yeh Shuhua chaired a small meeting on astrometry, attended by two more European astronomers, Paul Couteau (1923-2014) from Nice and Walter Fricke from Heidelberg, and by Dr. Hu Ningsheng (1934*) from Nanjing.

Dr. Hu was director of Nanjing Astronomical Instrument Factory (NAIF) and played an important role in our project. He visited me in Brorfelde the following year. In 1982 he organized a workshop at the Shaanxi Observatory on the Glass Meridian Circle with participants from five institutes in China, and I arrived from a stay in Japan. All my papers on the project had been translated into Chinese and printed. After presentations of Hipparcos and the GMC followed discussions where the director of the observatory concluded that the project should be continued as a collaboration on a novel technical development since we could not be sure of a success.

At home in Denmark, I made designs of the instrument which were converted to nice technical drawings in China. From May 1984 to June 1989 came many guests to Brorfelde from China, an astronomer

Dr. Li Zhigang (figure 13) from Shaanxi and several engineers, some of them stayed several years. My office was in Copenhagen, but I went to work with them in Brorfelde one day a week. They lived in the guest house where we then had Chinese lunch together.

In 1986, a plan for a Danish Chinese Meridian Telescope was agreed, to be built in the Nanjing factory and be set up at the Shaanxi Observatory. Professor Johannes Andersen (1943-2020) from Copenhagen visited Shaanxi in April 1986 to confirm the plan, and a Chinese delegation visited Denmark the year after, the president of Shaanxi Science Academy with two astronomers from Shaanxi Observatory.

I visited China several times to follow the development and give advice, and was always shown some attraction, e.g., Kunming Stone Forest, I climbed Mount Huashan in two days with four colleagues, I saw Guilin and Lijiang River, Shao Lin Temple, I saw the famous Longmen Grottoes near the ancient capital Luoyang, and my wife joined me on two journeys. I have seen more places in China than most Chinese and I have enjoyed talking Chinese with more people than most foreigners. In 1991 I visited again, but the project stopped after Hipparcos had become a success and had made fundamental astrometry with meridian circles obsolete. – I was happy to receive greetings recently and an invitation to speak online at the Shaanxi Observatory from Li Zhigang and his young colleague Yin Dongshan after they had read this report in draft.

Now I should come back to the astrometric problem mentioned at the beginning of this section, a systematic error of declination observations. The true cause was found during my work towards the new type of meridian circle. I think of this work with the greatest delight because the solution was so simple, some work of a skilled mechanic and hardware for less than a thousand Euros, yet it had marred astrometry for many decades, people had thought about it, and it took me twenty years to gradually track it down, to find the true cause and a simple way to eliminate the error.

Remains to explain the matter about tube flexure. The declination errors were not due to tube flexure but to tube refraction, i.e., refraction inside the telescope tube of the light passing from the telescope objective to the focus. This refraction was due to a vertical temperature gradient of about 1 C per meter but quite variable, Høg (1986). How this was found is detailed in Høg & Miller (1986) and Høg & Fabricius (1988). The gradient was removed by slow rotation of the air around the optical axis, and the simple arrangement of tangential air injection to maintain the rotation is shown in figure 11 of the latter paper. Figure 8 in the paper shows the resulting drastic improvement of the measurement of external refraction, i.e., the atmospheric refraction, a quite unexpected benefit, and this ends the story.

The referee wrote: "The understanding of the systematic errors caused by thermal stratification, and their successful removal merits insertion of the quoted Figures 11 and 8 of Høg & Fabricius (1988). It indeed was a major achievement and would have had a major impact in meridian astrometry was it not for the much better quantum leap caused by Hipparcos' success." - These two figures are now easy to find because I have included a link to the paper.

## 9. Hipparcos after 1980

After several years of scientific preparation, the Hipparcos space astrometry mission was eventually accepted by ESA's advisory committees and confirmed by ESA's Science Program Committee in Paris in September 1980. Before talking more about the work of the scientific teams, let me summarize, how these sorts of space science missions are organized.

ESA appoints a Project Manager (in this case Franco Emiliani) responsible for the technical development and a Project Scientist (in this case Michael Perryman, hereafter Michael) responsible for the scientific aspects, and ensuring that the science goals are achieved. Obviously, these two aspects must work closely together, respecting each other's constraints and goals.

ESA eventually selects an Industrial Consortium responsible for building the satellite. As Project Scientist for Hipparcos, Michael sets up a scientific advisory team to guide all aspects from start to finish. At the same time, ESA approved the setting up of three scientific consortia across Europe, one to compile the input catalogue that would be the starting point of the satellite observations and two independent teams to process the satellite data and create the final catalogue.

So, in 1981, we were sitting with a pile of technical notes which resulted from several years of theoretical studies about the mission's feasibility. I recall some of the first steps that ESA and the industry took in moving from these studies to a real mission.

What first comes to my mind is the invitation to a meeting in 1981 in Paris with engineers from MATRA, one of the companies competing to build the satellite. They wanted to gather an advisory team of scientists, which they did. We often met in Toulouse over the next few years. This was under a contract where we promised non-disclosure because there was a competing company also hoping to build the satellite.

ESA did not like such advisory teams very much, but they were tolerated. Now more on these contacts with the industry for the members of ESA's Science Team. Industry had significant 'catch-up' to do as they were newly exposed to a mission that was specified in quite fine detail as result of the Phase A study. So, they needed advisors, and so at the time ESA started the Phase B essentially all members of the Science Team for Phase A had accepted advisory functions.

Both Michael Perryman and on his advice also Vittorio Manno were rather unhappy at first, and they also strongly felt the need to keep the expertise of the Phase A Team in their Science Team.

Manno found the solution: in the coming time, the Science Team members could advice industry as they wished, but they should be careful to avoid foreseeable conflicts of interest between the ESA Project Team and the industrial Teams. In just one meeting it was agreed that the Science advisors should stay away from all advice that could lead to Industry trying to settle a dispute by saying: 'our advisors say…..' All Team members agreed and in fact such situations have hardly occurred and in almost all relevant cases the issue was already foreseen.

This setup has worked wonderfully well, also due to very careful dedication in this matter from Industry, hardly ever burdening the scientists involved, and it has generated a very collaborative atmosphere for the project!

Back to the first meeting with MATRA where I was the only scientist yet. They impressed me by showing several designs of the optical telescope system. Three of the designs were better than the one we had considered in our Phase A studies: I now understood how clever industry can be in optical design. In the end MATRA obtained the contract for Hipparcos, and we benefitted from their expertise and experience in many other areas as the project progressed.

## 10. The Hipparcos Science Team and pre-launch

From 1981 to the end of mission in 1997, the Science Team numbered about 12 people from across Europe covering different scientific, astronomical, and technical competences we met in person every 4-6 months, reviewing progress, and assigning tasks and priorities. I was one of ESA's scientific advisers throughout this time and I also led one of the two data processing teams, which was called NDAC.

About the Hipparcos Science Team - which Michael chaired for 17 years. What sort of topics were handled by the Science Team and what sort of preparation did this demand from my side?

I remember that a great deal of my time those years was dedicated to this work. My wife believes it was never out of my mind. We never took more than one- or two weeks summer vacation. Not until 1992 did I take as much as three weeks, enough then to bring us by train and ship as far north as Tromsø at the polar circle in Norway.

In preparation of the meetings, I read a lot of documents on technical or organizational matters, and we prepared documents for the Science Team to promote our ideas, and to present the mission at various meetings. I enjoyed the meetings, and I was always fully alert, while I sometimes saw a dreaming engineer when science was discussed.

I regret that I was unable to take good notes while a few of my colleagues wrote with a perfect hand in solidly bound books. I took notes but only on a piece of paper. I have sometimes asked Lennart Lindegren what happened at a certain meeting, and he just looked it up in his book. The chairman, Michael, always had a carefully planned agenda which was followed, and tables were always booked in a nice restaurant for the evening. I enjoyed these social occasions very much. I should add that we all shared the bill for the evening, but ESA paid our travel and hotel expenses.

Our Science Team meetings sometimes took place at the home institute of one of the team members. Such a meeting in Copenhagen had a session at the location where Ole Rømer put up the meridian circle he had constructed in 1704. This place was 20 km outside Copenhagen because Rømer's official observatory on top of the famous Round Tower was considered unsuited for his delicate observations of positions.

Most of our meetings were at ESA's Research & Technology Centre at Noordwijk in The Netherlands. There was always a lot of technical information provided by the project team and guidance requested from the scientific advisors. There was a nice speed of progress all the time after 1981.

There is a big gulf between early ideas for a space science mission and the detailed design, the technical development, the integration and testing, and of course the launch and operation. After the Phase A studies in 1980, which led to a baseline mission concept there were no major new technical developments made to the basic design. The baseline was maintained, but it was now implemented in reality, e.g., with a good optical design as I mentioned. Gas jets were introduced for attitude control instead of the reaction wheels that I had imagined.

During the Phase A Studies I made several innovative contributions to the instrument and satellite design. But in the period, we are talking of now, no aspects of the instrument, or the satellite technology, raised my concern. I only remember that I always felt I was doing useful work on a great astronomical project. Let me focus here on some of the satellite data analysis aspects.

Before launch, from 1981-1989, our consortium NDAC (for Northern Data Analysis Consortium) was preparing for the satellite data. There were two data analysis teams who developed the software independently and would process the satellite data independently - the other was the FAST Consortium, led by Jean Kovalevsky. We called ours the northern consortium because the FAST consortium was more in the southern part of Europe.

From 1981 we had to build a consortium as required by ESA. It was clear from the beginning that it should involve the Royal Greenwich Observatory where I had long contact with Andrew Murray - whom I had brought into ESA's Astronomy Working Group (AWG) to succeed me. Also, Mike Cruise from Mullard Space Science Laboratory was supporting the project. I had always seen Murray as the natural consortium leader, but he insisted that it should be me when we began forming the consortium called NDAC.  Of course, Lund Observatory with Lennart Lindegren should come in. In Copenhagen I had support from the Danish Space Research Institute, and from the Danish Geodetic Institute. In both cases we could get access to computing facilities and to methods of treating large systems of equations, especially through Knud Poder (1925-2019) from the Geodetic Institute.

When we were building up the data processing software for treatment of the real data there were many studies and simulations undertaken, and the results of these simulations and tests were fed back to the design of the satellite.
We undertook the first simulations of data processing in Copenhagen in 1977-79, and thay were crucial for the design of angular one-dimensional observations with an astrometry satellite with revolving scanning. We could show that the subsequent reconstitution of the celestial sphere could give (1) relative positions and (2) absolute trigonometric parallaxes, for a network of stars covering the entire sky, after only one year of observation (Høyer et al. 1981).  Many other simulations were made to develop the data reduction chain of NDAC, and these studies led to the choice of basic angle, piloting of the detector, gas jet firings for attitude, etc.
Let me mention the main task leaders of the NDAC consortium: these people stayed with the project for its entire duration! Many people were involved in the consortium: all 19 are listed by name in the final catalogue. Let me mention here those working on the data reduction during most of their time: Carsten Skovmand Petersen and myself from Denmark, Lennart Lindegren (Consortia Leader from 1990) and Staffan Söderhjelm from Sweden, with Dafydd Wyn Evans, Floor van Leeuwen and Andrew Murray from the Royal Greenwich Observatory, United Kingdom.
I always aimed at a collaboration of very few institutes: at RGO, Copenhagen and Lund for the sake of efficiency and ease of organization, but efficiency was also a hard necessity with the very limited resources I could expect to have.
Our approach to the data analysis problem was quite different from the structure of the other data analysis team, called FAST. Two consortia were working in parallel on the Hipparcos observations to compare the results from two independent groups for verification of these unique space observations. Jean Kovalevsky led the other consortium (FAST) and he created a much bigger organization than we did for NDAC. He wanted to involve as many countries in Europe as possible to make Hipparcos a real European mission, a very laudable goal indeed, and he succeeded with a consortium of 82 persons in five countries: France, Germany, Italy, The Netherlands and (a few from) the US.  In this way, Kovalevsky paved the way for further European space astrometry after Hipparcos.
The limitations that existed in computers and computer networks at this time meant that the successive steps of the data analysis were made by physically posting the data from one institute to the next on big reels of magnetic tape each containing a "staggering" 100 megabytes.

In NDAC the satellite raw data were first treated at the Royal Greenwich Observatory (UK), and the output was sent on magnetic tapes to Copenhagen, where the abscissae of the stars along great circles were derived. They were used to derive the five astrometric parameters for all stars. At Lund, in Sweden, the processes were defined mathematically, and the double stars treatment also took place there, led by Staffan Sõderhjelm and Lennart Lindegren.

## 11. Tycho

When Hipparcos was accepted by ESA in 1980, it targeted measuring 100,000 stars to a typical accuracy of 2 milliarcsec, 2 thousandths of an arcsec, we ended up doing "a bit better" than that in practice: almost 120,000 stars to an accuracy of 1 milliarcsec, the factor of 2 is considered a quite significant improvement of accuracy among astrometrists.

But a big development came in 1981, with my ideas for an add-on to the satellite, which I called Tycho. Let me quote, with recent permission, from Michael's account of the mission which gives the background:

"In April 1981, already well into the satellite design phase when significant modifications would normally have been rejected out of hand for reasons of risk, and for the increased cost that they could incur, Erik Høg pointed out that a rich stream of satellite data was lying untapped. He realized that the signals from the satellite attitude detectors contained a staggering quantity of star positions that were, quite simply, not being sent to the ground. But once pointed out the harvest was obvious: positions for the million or more stars not being observed by its main detectors. Working quickly to optimize the system, and with a little urgent lobbying, the small amount of additional funding needed was made available. A color filter and additional detectors were added, an extra telemetry channel set up to send the data to ground, and a new data processing team put in place to handle the data stream".

So, in addition to the two existing processing teams we set up another - the Tycho Data Analysis Consortium (TDAC) which I then led through to completion. This idea of using the satellite's star mappers to get this extra star catalogue came to me as follows.

The idea came in March 1981 when I wanted to define suitable meridian circle observations of reference stars for the Hipparcos attitude determination. I realized that the photon counts from the star mappers (figure 12) could be used to detect many more stars than the few thousands required for the attitude. My estimate was that positions and magnitudes could be obtained for at least 400,000 stars, and perhaps many more. But this required that all data had to be transmitted to ground, about 100 Gbytes, the same amount as for the main Hipparcos mission. I immediately wrote three short notes to ESA about this *Tycho project* as I called it.

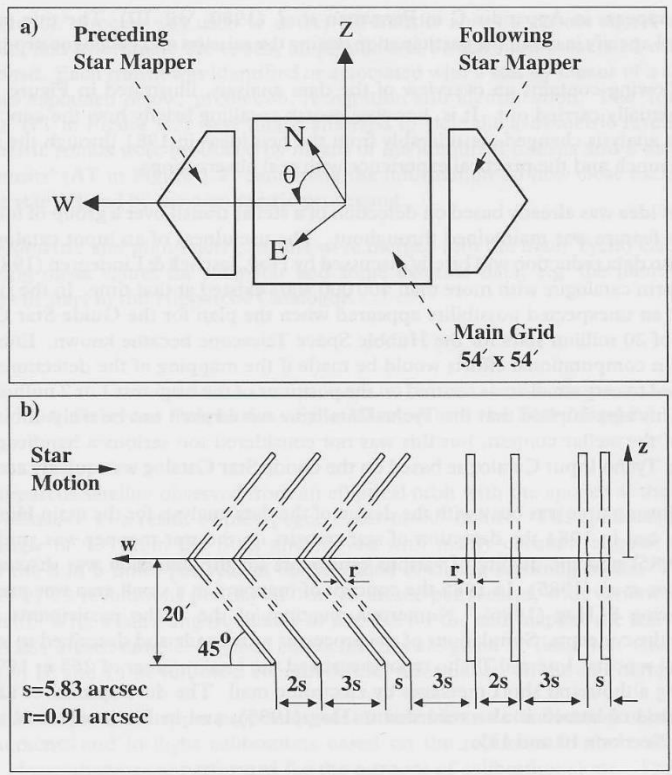

*Figure 12. The slit systems at the focal plane of Hipparcos. (a) Arrangements of the star mapper and the main grid. (b) The preceding star mapper; the following star mapper is redundant and was in fact never used. The 'vertical' slits of the star mapper are perpendicular to the motion of the stars, while the 'chevron' slits are inclined by 45º. The light from the whole star mapper area is divided by a dichroic beam splitter onto two photomultiplier tubes which count the photoelectrons simultaneously in two bands: $B_T$ and $V_T$, close to the traditional B, blue, and V, visual, photometric bands in astronomy. This star mapper is the Tycho experiment which obtained both astrometry and two-color photometry on the Hipparcos satellite. - Figure from ESA (1997) Vol. 4.*

Michael continued: "This was really a very interesting example of something which was quite obvious to everyone as soon as you pointed it out but had escaped everyone's attention except your own. But I think that everyone on the Science Team immediately understood its importance?"

The proposal was indeed well received by Michael, and in the Science Team. I remember especially seeing the faces of Michael Grewing and Jean Kovalevsky at the meeting. The name Tycho was accepted without discussion, and this was a great relief for me since I remembered that Kovalevsky had introduced the name Hipparcos for the satellite instead of the name Tycho that I had wanted.

Of course, convincing ESA to find and approve the funding for the change s was another matter! Later in 1981, ESA approved the implementation and the extra cost. From Vol.4 XI of ESA (1997) we read: "Had the Tycho proposal come a couple of months later, the satellite design would have been frozen, and the idea of the Tycho project would have been a lost opportunity. This wonderful idea would have been difficult to forget about, even though we were immersed in the fascination of the main mission, and in all the work it gave us."

The data reduction for the Tycho experiment still had to be set up and organized. So late on in the mission, and with a such a big data volume, that was not an easy challenge at all. I recall a moment at the beginning when I was overwhelmed having to do that and said so to Lennart Lindegren, when we were waiting for the bus somewhere. He just smiled and said he was sure we could do it, I then even believed he would do it in Lund! Lennart was always a great support. I never hesitated, when I had the good ideas in those years, and luckily, I did not know how big the tasks would be.

Michael Grewing was the first to tell his interest by letter when I proposed the Tycho project in 1981, and as director of the astronomical institute in Tübingen he maintained an excellent young team there over the years. Carlos Jaschek (1926-1999) was director of the "Centre de Donneés Astronomique de Strasbourg" (CDS) and invited the Tycho project. The CDS is hosted by the Observatoire Astronomique de Strasbourg, now known by its acronym ObAS and I remember Daniel Egret and Michel Crézé as directors there during the Tycho years. A team at Strasbourg did a wonderful job. But these two teams, together with Copenhagen and Lund, were still not enough for the task ahead.

Derek McNally (1934-2020) in the UK tried to form a team, but continued support for a team failed. So, we had a big problem. But then Roland Wielen as director of the Astronomisches Rechen-Institut in Heidelberg, who had recently taken over from Walter Fricke (1915-1988), offered to establish a team and of course I gladly accepted. Wielen recently wrote: "I think you describe the start of the ARI contribution to Tycho properly. ... I thought that the Tycho project was very promising and should be supported by the ARI. Hence, I proposed early in 1986 that ARI should become a member of TDAC. I think it was the right decision."

Uli Bastian of that team told me later that at first, he was reluctant to work on Tycho because he wanted to concentrate on the PPM Star Catalogue, the production of which he had just started together with Siegfried Röser. But he also saw that it was a correct decision by Wielen, and so he took up the task - and he did it wholeheartedly throughout the years.

All this led to the Tycho Catalog of a million stars, published at the same time as the main catalogue in ESA (1997). A scientific team was organized as the Tycho Data Analysis Consortium, TDAC, to process this additional data.

The observations consisted of photon counts from the satellite's star mapper slits, continuously obtained during the 37 months of mission. Detection of a star in this data stream - or rather of a signal above the noise which could therefore be a star crossing the slits - that was recorded as a function of time. The first paper on Tycho detection was by a guest in Copenhagen from Tokyo Observatory, Masanori Yoshizawa (figure 13), (Yoshizawa, Andreasen, Høg 1985).

Stars were later confirmed and identified with real stars in the sky using two sources of information: 1) accurate knowledge of the pointing of the satellite telescope for every moment of time, the so-called satellite attitude, and 2) a list of 3 million stars with positions known from ground-based observations.

This may seem simple in principle, but it was quite complicated in practice, because the work had to be done within a tight time schedule and it had to be divided between three institutes due to the limited computer capacity and manpower. It is fair to say that no astrometric research before had been done under such constraints as the Hipparcos and Tycho data reductions. Therefore, we owe a cordial thank you to all those working in the consortia for their dedication to our great projects during many years.

In addition to new people brought in to the Tycho Consortium, many individuals already working in the NDAC and FAST teams also contributed expertise and data. The Tycho Input Catalogue of 3 million stars was based on early access to the Guide Star Catalog for the Hubble Space Telescope as provided by Brian McLean and Jane Russell. The satellite attitude was provided - from the main satellite data processing - by Franceso Donati and Pier Luigi Bernacca in Italy (from the FAST Consortium) and by Floor van Leeuwen in the UK (from the NDAC Consortium).

Many people were involved in the Tycho Consortium: all 37 are listed by name in Vol. 1 of the final catalogue ESA (1997). Let me mention those working on the data reduction during most of the time: Claus Fabricius, myself, Valeri Makarov, Holger Pedersen, and Carsten Skovmand Petersen from Denmark, Daniel Egret, Jean-Louis Halbwachs, Jean Kovalevsky, Catherine Turon, and Pierre Didelon from France. Further members were: ten from Germany, two from Italy, two from The Netherlands, one from Sweden, one from Switzerland, four from United Kingdom, and two from the USA.

We had many meetings of the Tycho Consortium during the 1980s at the participating institutes, but one of the meetings stands out in particularly in my memory. It took place in 1984 on the island Hven, (see figure 13) the historic site near Copenhagen where Tycho Brahe with his many collaborators made his famous observations of one thousand stars and of planets during twenty years from 1576-96. At the meetings we discussed the progress of work and defined how to proceed - and we had a nice and productive time together. I stayed a year as guest professor at the institute in Tübingen, two months in Tokyo, and two months in Strasbourg.

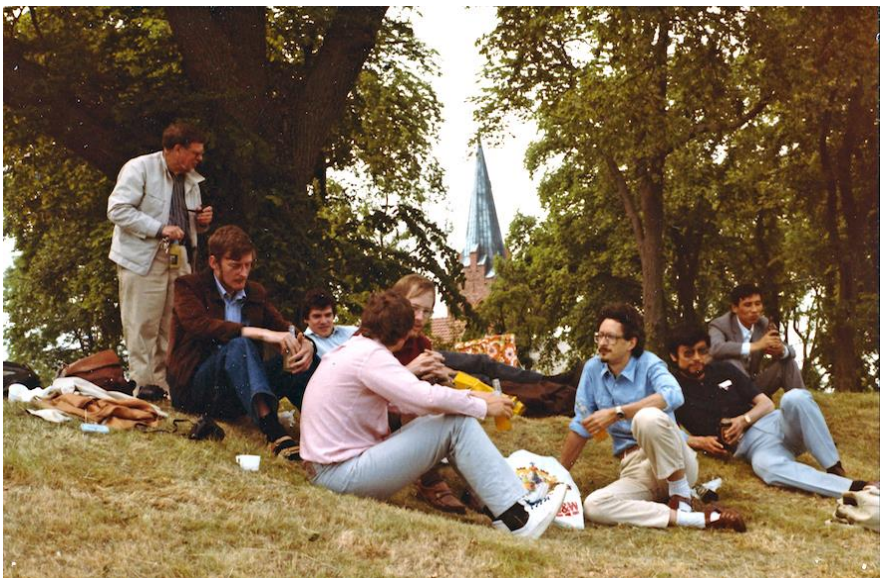

*Figure 13. The Tycho Consortium in June 1984 at lunch on Hven after our discussion of the Tycho experiment on the Hipparcos satellite. We were sitting precisely where Tycho Brahe's renaissance palace Uraniborg had been standing. This spot was in the year 1984 just a grass covered hole in the ground as it had been since the peasants took the stones for their own houses soon after Tycho Brahe had left the island on 29th April 1597. From left: Knud Poder, Georg Kjærgaard Andreasen, Michael Perryman, Lennart Lindegren, Carsten Skovmand Petersen, and Staffan Söderhjelm. Guests from Japan and China: Masanori Yoshizawa, and Li Zhigang. - Figure from the private collection of Lennart Lindegren.*

We mentioned that the Tycho Catalogue was published in 1997, at the same time as the main Hipparcos catalogue. But this was all followed by the improved Tycho-2 Catalogue in 2000. One big difference for our work on Tycho-2 beginning about 1997 was that we collected all the star mapper photon counts from the whole mission close to any star in the input catalogue. From the resulting fifty or more detections we could decide whether we had a real star with much better statistical certainty than from the single detections used for Tycho-1 and this resulted in accurate positions for 2.5 million stars, a fantastic number for that time. Crucial for this result and for the better accuracy for all stars was that Valeri and Claus applied a novel method of data processing suitable for faint stars.

The possibility to combine all photon counts had been clear to me all the time. During my one-year stay in Tübingen 1987-88 we had many discussions and Andreas Wicenec started some simulations of the data reductions. But it had only become realistic with the greater computing power we could buy ten years later. We were also able to get the needed private funding for the computer and for Claus Fabricius and Valeri Makarov to do the work in Copenhagen. Andreas Wicenec supplied the satellite data from Tübingen, all together 3000 nine-track tapes of 1 kg each with 300 GB were condensed to 45 DLT tapes, a new kind of smaller magnetic tapes. Uli Bastian and Peter Schwekendiek in Heidelberg supported with computations, and the final satellite attitude was provided by Floor van Leeuwen at the Royal Greenwich Observatory.

And then, the other big advantage over Tycho-1 is that Tycho-2 contains accurate proper motions. They were based on the Tycho-2 positions, which were combined with a new analysis of the positions observed during the past 100 years and given in 144 ground-based astrometric catalogues provided by the US Naval Observatory in Washington DC by Sean Urban.

I recall meeting Sean at a conference in 1998 in Gotha and realizing this fantastic possibility. They had not yet completed the new analysis, but we received all the positions just in time to derive the proper motions for our publication in 2000. Had they come a bit later, our funding would have ended before we could do so. Tycho-2 has been the most cited astrometric catalogue ever since 2000 apart from the Hipparcos Catalogue.

Close double and multiple stars could also be treated much better in the Tycho-2 processing. The discovery of 13 251 candidates of visual double stars, mostly with separations between 0.3 and 1 arcsec, is reported by Fabricius et al. (2002). Claus Fabricius and Valeri V. Makarov published 2-colour photometry for almost 10,000 components in systems with separations above 0.3 arcsec.

The Tycho-2 Catalogue took on a renewed importance with the advent of Gaia, because the positions from Tycho-2 from the early 1990s could be combined with the early Gaia positions to give the TGAS (or Tycho-Gaia) Astrometric Catalogue with almost 25 years of proper motions and which was also important for the early Gaia Astrometric Iterative Solution. But I have not been involved in this latest TGAS Catalogue, except sharing the joy over TGAS. The rate of citation for Tycho-2 has stayed quite constant even after publication of Gaia results beginning in 2016. Even today, the Simbad database gives the photometry (B, V) deduced from Tycho-2 as 'fundamental data' for many stars.

## 12. Launch of Hipparcos

Hipparcos, figure 14, was launched by Ariane 4 from French Guyana in 1989. As one of the key scientists, and one of the consortium leaders, I attended the launch. It was a great experience, seeing the launch, and witnessing this new chapter in astrometry.

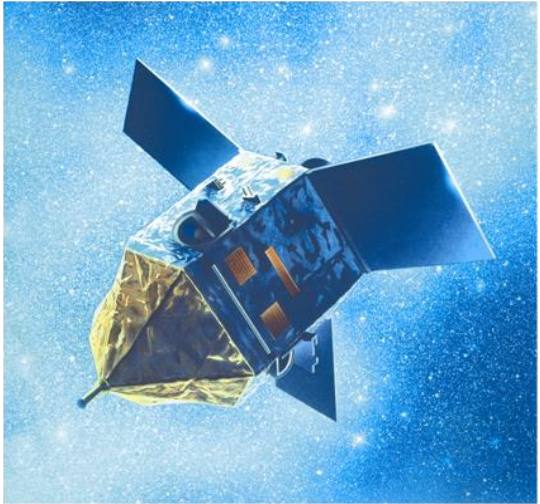

*Figure 14. The Hipparcos astrometry satellite, launched by ESA in August 1989, a revolutionary event in the history of astrometry and astronomy. - Figure from the private collection of the author.*

The Hipparcos Science Team and many others witnessed the launch in Kourou, but some of us had the special honor to fly in the super-sonic Concorde. From Hipparcos, we were the consortia leaders as well as Professor Pierre Lacroute, the father of space astrometry. We flew from Paris to Dakar and then on to Kourou, far above the highest clouds, and we could see the setting Sun, all the time at the same angle above the horizon while we crossed the Atlantic. I had a window seat and noticed that the wall became warm during flight, due to air friction at the super-sonic speed.

We saw the impressive launch facilities in Kourou, had our luxurious breakfast at the beach, and we witnessed the launch from a spot very close to the rocket. We saw the fire and smoke and we saw the big rocket slowly lifting off. But we heard nothing, not until many seconds after. We were happy seeing the glowing fire from the rocket moving away up there, and we toasted in champagne.

After a spectacular and flawless launch, we had big problems - the apogee boost motor intended to circularize the satellite orbit failed to operate and the satellite was stuck in an elliptical 10-hour orbit rather than the planned geostationary! These events had a tremendous impact and all but killed the whole mission. Only very hard and dedicated work from the engineers and the new Project Manager (Michael Perryman) succeeded in overriding the need for a new mission that we were still thinking and negotiating about through the end of 1989. Yet leaving the mission in the Geostationary Transfer Orbit (GTO) required colossal and unforeseen adaptations.

The launch itself may be called flawless. That it was indeed, but only after a very scary hick-up. At the end of the first countdown nothing happened.....!!

At the time a terrible scare!

It soon appeared to be due to a stop at $t_0 - 7$ seconds, which fortunately was 1 second before the detachment of the cryo-arms that keep the rocket filled up. Otherwise, the launch, which now took place already August 8th, 1989, well into the regular holidays of the launch crew would have had to wait till after the holidays. With the cryo-arms still in place a search for the cause of the interruption could successfully be done in about 1 hour including the (software)reset, and a countdown restart at $t_0 - 15$ minutes could still be setup with 5 minutes margin to the end of the launch window.

This time we attended a flawless launch.

Perhaps good to recall the 10 hours elliptical orbit as result of the launch into the GTO (Geostationary Transfer Orbit), and to point out that the orbit had to at least execute a perigee rise. That completed, the satellite still had to face a devastating occultation period (earth shadow) around Easter 1990. The new unforeseen orbit had significant impact on observing modes (basically everything foreseen had to be adapted) and so it finally took until mid-December 1989 to successfully arrive at the first position solutions with the target accuracy.

A colossal joy at the Science Team meeting in Darmstadt, when Hans Schrijver (working still on the first-look) called he had reached below 0.1 arcseconds rms error. First proof of principle!

Until the occultation date the Project Team and Michael Perryman in the new role of Project Manager kept finding more and more ways to minimize power consumption, and in fact the mission passed through the biggest occultation with a few percent leftover in the batteries, having stopped much of the hardware that was not even designed to turn off and on again. So, by summer 1990 the mission finally was in a stable observing mode and condition, be it that the radiation dose received was 2 orders of magnitude more than budgeted in this unforeseen orbit environment. This finally killed the satellite in March 1993 shortly prior to it running out of attitude control gas.

I don't want to dwell further on the enormous technical problems that followed but instead let's concentrate more on the treatment of the satellite data. I do not remember the moment when the first "reference great circles" were constructed and when the first sphere solutions or first Tycho results started to appear. But I remember the IAU Symposium No. 141 on 16-23 October 1989 of the Pulkovo Observatory near Leningrad (now St Petersburg) where I had been asked to present Hipparcos on behalf of the Science Team. It was during the months after the post-launch apogee boost motor failure, and everybody asked about the latest news from the great Hipparcos mission - which had a very uncertain future after the boost motor failure. A resolution was adopted by the symposium about Hipparcos stressing the importance of the mission, congratulating the ESA teams for their efforts to operate the satellite in the elliptical orbit, and recommending a second mission if need should be.

In the presentation, I had shown a picture of the first Tycho observation. It was a plot with the very first recording of transits over the four slits in the star mapper during satellite commissioning. I had made several prints for distribution, and I can still see all the eager hands reaching to get such a print of the first Tycho observation.

A distinguished person approached me during a break. Away from the crowd he presented himself as coming from the Russian space agency and offered to launch a second Hipparcos in case ESA could not.

And this is where I first met Valeri Makarov, who played a big part in the Tycho processing. Viktor K. Abalakin (1930-2018), director of the Pulkovo Observatory, introduced me to his young people: Mark Chubey (1940-2016), Vladimir Yershov, and Valeri V. Makarov. I had before heard of Russian plans to launch a Hipparcos successor ten years later to derive very accurate proper motions from positions at the two epochs. Now I met these people with whom I had very fruitful discussions during the following years. Already the following year when I visited Pulkovo again, we began developing a Hipparcos successor which became Gaia as I will report in section 14, and Valeri Makarov soon became my very bright collaborator in Copenhagen.

I also recall a conference in Cambridge probably in 1990 when the first analysis of some Tycho observations had been made. I could then see that we would be able to make a catalogue of one million stars, not only 400 000. That was great news. Two of the reasons were a slightly higher quantum efficiency of the photomultipliers, and a more accurate background determined with a method invented by Andreas Wicenec using the median value of the photon counts during an interval of time. A third reason was the development of a de-censoring algorithm that became even more essential as the background noise varied due to van Allen belt crossings.

I think it's worth emphasizing that the final catalogue accuracies, both for the main Hipparcos Catalogue, Perryman et al. (1997) and ESA (1997), and for the Tycho Catalogue were not just close to the theoretical predictions that had been made much earlier on in the mission design and at the completion of the phase A study. - for such a complex instrument, this would already be a great achievement. No, the predicted performance was significantly exceeded in several respects.

The final accuracy of Hipparcos reached 1 milli-arcsecond instead of the predicted 2 milli-arcsecond, and this is quite significant in the view of an astrometrist. And the Tycho-1 Catalogue contained one million stars - compared to the 400 000 in my original proposal. The Tycho-2 has even 2.5 million stars and it contains accurate proper motions, and it includes two-color photometry for every star. So Hipparcos performed much better than expected despite the bad orbit. We really must admire the responsible operations team for their technical expertise; you ought to read the exciting story in chapter 8 of Perryman (2010) about the mission recovery after the boost motor failure. The data reduction consortia had to adapt to the new conditions, in Tübingen, e.g., essentially the whole detection pipeline for Tycho had to be re-written.

### 13. Finalizing the catalogue

The eight years from 1989, when the satellite was launched, to 1997 when the final results were published, were very hectic. All the teams working to a tight schedule, to get the results out to the wider astronomical community, when research based on the data was started.

I had handed over leadership of the NDAC Consortium to Lennart Lindegren soon after launch to focus my efforts on the Tycho data, and I presented the Tycho Catalogue for the first time, at a major international conference in Venice in 1997. The Venice conference was a great event in every respect, the setting itself in Italy where I knew that Michael and Pier Luigi Bernacca (1940-2013) had done the perfect preparations. The results we could present were very well received by the wider astronomical community. The whole story of Hipparcos has been told in the book by Perryman (2010), including the highlights and the problems, even disasters on the way.

I gave many lectures and interviews with the press, after the catalogue publication. One event stands out especially, the big International Astronomical Union conference in 1997 in Kyoto, Japan, where we could present our results - and we could again see they were very well received.

Michael has asked: "Looking back now Erik, in one lifetime you have helped take astronomy from the confines of the Earth out into space with two highly successful space missions. You could never have imagined such progress when you started out in the field 60 years ago?"

No, I never imagined anything like that. I began with astrometry at the new meridian circle in Denmark in 1953 when I was just 21 years old. In Copenhagen I saw that astrometry was central for astronomy. But soon after I understood that astrometry was not so highly rated everywhere. From 1958 at the Hamburg Observatory at first on a 10-month fellowship, I wanted to do astrophysics because I clearly saw that real astronomy happens in that area while astrometry was rather old-fashioned.

But in July 1960 something happened. I had a revelation you might say, the vision of photon counting astrometry with meridian circles. It could be done using computers, punched tape, and all that. That brought me back to astrometry. It seems after all, that I have done more for astrophysics than I could ever have done as an astrophysicist.

The idea was well received in Hamburg-Bergedorf, the director Otto Heckmann saw that it should be used for the meridian circle expedition to Australia which was in preparation. And Heckmann trusted me, he trusted that I had the talent and that I had the stamina to hold on to the project for years. I had the luck to have the right idea at the right place and right time and I had the luck to meet the right people. The same luck I had later with the design of Hipparcos in 1975, the Tycho star mapper project, and the Roemer mission with CCDs which led to Gaia. It seems that the luck was there again with the proposal in 2013 of a Gaia successor in twenty years, a project now having a high ESA priority and a quite probable launch in 2045, Høg (2021).

## 14.  The billion-star astrometry after 1990

What were the steps taken in advancing space astrometry after Hipparcos? There was excitement and interest around the world to extend space astrometry to more stars and better accuracy. And there were many ideas.

In the US there were (POINTS, USNO - NEWCOMB, MAPS, FAME, AMEX, OBSS, JMAPS, OSI; by leading astrometrists, including Reasenberg, Duncombe, Jefferys, Seidelmann, Johnston, Hemenway +) and in USSR/Russia (Lomonossov, REGATTA-ASTRO, AIST, LIDA, OSIRIS; by Chubey, Makarov, Yershov, Nesterov +). But even SIM (by Mike Shao) which did very well in the US Decadal Survey in 2000, was abandoned in 2010. DIVA (in Germany; by Siegfried Röser, Uli Bastian +) targeted 0.2 mas to 15 mag, JASMINE[1] (in Japan by Gouda, Yamada +) was still being developed, STEP (exoplanet detection) was proposed in 2013, and there were also other astrometric missions proposed to ESA: NEAT (M3) and Theia (M5).

A total of 18 names of ideas or projects are listed here. But all of these - except JASMINE scheduled for launch in 2028 - eventually fell by the wayside leaving Europe, again, in the "driving seat" in space astrometry.

There are several reasons. The ideas needed for space astrometry started in Europe already in 1960 with photon counting astrometry, and in 1975 the basic realistic design of a scanning satellite was on the table – as reported above. A year before that, in 1974 the early ideas for space astrometry had changed from being a purely French undertaking to become a European project within ESA. That opened a major opportunity for the still existing astrometric community in Europe to become active, and so they did.

Things then moved quite quickly from these early ideas in Europe, to reality. Given the basic design and the active scientists, European industry was able to develop the Hipparcos satellite for approval in 1980 and launch in 1989. The European astronomical communities in ESA countries joined efforts to treat the observations and produce the final Hipparcos and Tycho Catalogues in 1997, and they surpassed all expectations with respect to accuracy and number of stars. All this gave a solid foundation in ESA, in European industry and in the scientific community for future space astrometry in Europe when new good ideas appeared.

---

[1] Yoshiyuki Yamada has informed me about JASMINE: The current launch schedule is officially 2028 and he continued: "We are now collaborating with ARI group for the JASMINE data reduction. ARI Heidelberg Group and our group will handle the data independently. This is like the relationship between FAST and NDAC for the Hipparcos data reduction."

It is characteristic that the development in Western Europe has followed a very straight path: From meridian circles to Hipparcos, then to Gaia, and now towards GaiaNIR which will probably be launched about 2045 as a large ESA mission. That is an interesting point about the common development of the scientific and industrial aspects given this enormous basis for future ideas.

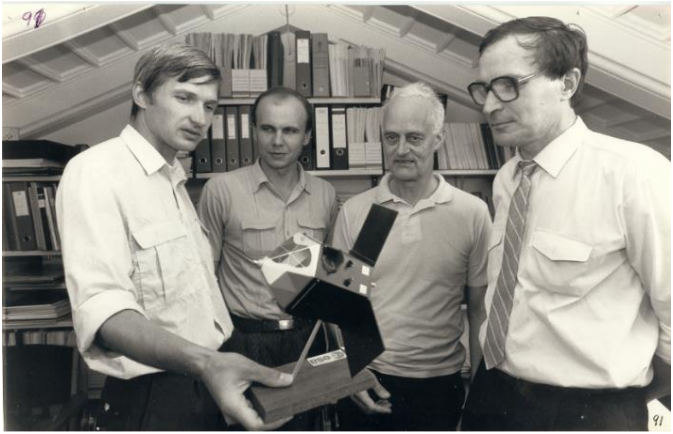

*Figure 15. The Russian space astrometry team visited Copenhagen in 1991 for discussion of the ideas of an Hipparcos successor which resulted in the Roemer proposal in 1992 and eventually became Gaia launched by ESA in 2013. In my office at the Observatory on Øster Voldgade, we are looking at the Hipparcos model, V. Yershov, V. Makarov, E. Høg, and M. Chubey leader of AIST, a planned Hipparcos successor to be launched in Russia. One of the members, Valeri Makarov, then stayed seven years in Copenhagen working on the Tycho-1 and Tycho-2 Catalogues. - Figure from the private collection of the author.*

The first steps along the path to a new space mission in Europe after Hipparcos was taken already in 1990. During a visit to the Pulkovo Observatory I discussed with my USSR colleagues their existing proposals for a successor to Hipparcos. After a day trying to understand their design, I realized that I could not understand how it would work, and that instead I was in the process of designing a better more accurate Hipparcos satellite. They took me to the space center near Moscow and from there to the Pulkovo observing station in the Caucasus Mountains, at Kislovodsk. In the first night there I woke up and made the first design drawing on paper. Mark Chubey heard me and thought I was ill, but then he understood and brought me a better lamp. In the morning we woke up to a view I won't forget, the clouds were gone, and we saw the majestic snow covered twin peaked Mount Elbrus only 65 km away.

This generated further visits to Pulkovo, Moscow, Copenhagen, and Lund for discussion of the ideas with Lennart Lindegren, Mark Chubey, Valeri Makarov, and Vladimir Yershov, figure 15. Finally in 1992, I made a design with Lennart Lindegren of a mission with CCD detectors which we called Roemer, and which later developed into Gaia (Høg 2001c). Roemer was first presented at a Symposium of the International Astronomical Union in Shanghai in September 1992, Høg (1993).

All this was going on soon after Hipparcos had been launched, well before the final Hipparcos Catalogue had been published. Main features of this design differed from Hipparcos. The proposal used direct imaging on Charge Coupled Devices, CCDs, with time-delayed integration, and it contained many other features implemented in the final Gaia, figure 16. For a 5-year mission an astrometric accuracy of 0.1 milliarcsec was predicted at 12 mag, more than 10 times better than Hipparcos. The astrometric efficiency was 100 000 times higher, because CCDs have ten times higher quantum efficiency than photomultipliers, and because thousands of stars could be observed simultaneously, and this was obtained with the same telescope aperture of 29 cm as Hipparcos. Astrometry and multicolour photometry for 400 million stars were also included.

A comparison of Hipparcos with the final Gaia satellite: Almost $10^9$ CCD pixels integrating simultaneously with almost 5 times the DQE, detection quantum efficiency, in comparison to the photocathode of the IDT, and with an integrating TDI mode on the CCD to suppress the Read Noise. This should be compared to a single IDT integrating only one star at a time. - Optically we finally arrived at a much larger entrance pupil in the along-scan direction, giving some ~$6^2$ times larger astrometric weight per detected photon as compared to the small aperture, not rectangular, Hipparcos pupil.

In fact, together with the much longer mission duration this explains the $10^7$ factor weight increase of the Gaia mission over Hipparcos.

### Focal plane with CCDs

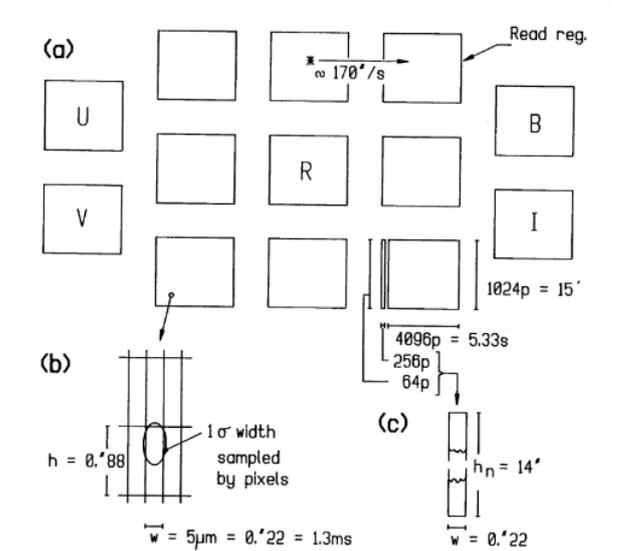

### Sampling

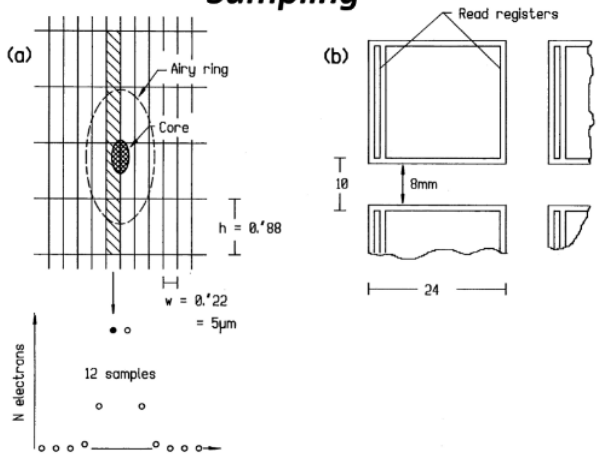

Figure 16. Design of Roemer in 1992, proposal for an astrometry satellite with CCD detectors in the focal plane, a design which became basis for Gaia. Below left: The sampling of data from the CCDs with pixels is matching the diffraction image from the telescope. Below right at (b): Note that the short CCD for bright stars is followed by a long CCD giving longer integration time as needed by faint stars. The stars move from left to right as the satellite is slowly spinning, the charges in the pixels are moved at the same rate and are read out when arriving at the read register. This is called TDI, Time Delayed Integration. - Figure from Høg (1993).

Going on in ESA at this time was the long-term planning for European space science, called Horizon 2000, which had been initiated by ESA's Director of Science, Roger-Maurice Bonnet. We saw the next opportunity for space astrometry with ESA's call for a new mission in 1993. The Roemer proposal had aroused much interest in the Hipparcos Science Team. This Team supervised all aspects of the Hipparcos mission until its catalogue publication in 1997, and it was still the main forum for top-level discussions amongst the European teams involved in space astrometry.

The ideas in Roemer were adopted in a mission proposal submitted to ESA on 24 May 1993 for the Third Medium Size ESA Mission (M3), Bastian et al. (1993). The proposers from seven countries were: Kovalevsky, Lindegren, Halbwachs, Makarov, Høg, van Leeuwen, Knude, Bastian, Gilmore, Labeyrie, Pel, Schrijver, Stabell, and Thejl. We proposed to measure 100 million stars and to obtain an accuracy of 0.2 mas at V=13 in a 2.5-yr mission with a 34 cm telescopic aperture. At the faintest magnitudes 15-17, accuracies would be about 1 or 2 milliarcsec. i.e., a similar accuracy to Hipparcos but much fainter.

This mission was not one aimed at solving some particular scientific problem, but rather to explore the rich complexity of Nature in order to find new patterns, anomalies and even contradictions to existing ideas. This was the spirit of the proposed Roemer mission.

It was very broad in its reach. But we did give some details about the scientific importance and objectives. They were detailed in ten pages of the proposal, covering stellar structure and evolution, stellar kinematics, cosmic distance scale, double stars and unseen companions, solar system objects, general relativity, and the celestial reference frame. The proposal, compiled and edited by Lennart Lindegren, had 28 pages. And there were two annexes, which I edited running to more than 60 pages.

An acronym was used: ROEMER for "Rotating Optical Observatory for Extreme Measuring Efficiency and Rigor". A section called "The FIZEAU option" was included "to point out a possible improvement towards a scanning satellite with ten times the angular accuracy of ROEMER". The section described a larger scanning satellite with "two confocal Fizeau-type (or 'wide field') interferometers whose axes form a basic angle of the order 140 deg.". This sticks again to the measurement technique involving two viewing directions with an extremely stable 'basic angle' between them as a key idea. It was a description fitting very well to the later GAIA but the included optical system underwent major development before it was called GAIA, an acronym for Global Astrometric Interferometer for Astrophysics.

Roemer, GAIA, and Gaia all stick to two other key ideas introduced by me in December 1975 (see section 7) the *revolving* and *one-dimensional* scanning. In January 1976 I introduced a further key idea: *very careful choice of the basic angle of the beam combiner. These four key ideas about the beam combiner and scanning* are basic for the European lead in Space Astrometry, which is always global astrometry, all-sky astrometry.

The Roemer proposal was submitted to ESA in May 1993, and was then reviewed by ESA's Astronomy Working Group. To our delight, our proposal was rated the best scientifically among all astronomical proposals for M3. But the bad news was that it was considered to come too soon after Hipparcos and it was not sufficiently ambitious with respect to accuracy. It was therefore referred to a Cornerstone Mission study if 10 - 20 μarcsecond accuracy could be demonstrated.

In 1993 Roemer really "lost out" to the cosmic microwave background mission Cobras/Samba later called Planck. At that time, I did not think of the competing and winning mission, only that we had lost. So, the Astronomy Working Group considered Roemer's science very important for Europe, but that the mission was not "ambitious enough".

That is what I remember. I thought it was a severe blow to a good proposal and that a ten times better accuracy was rather unrealistic. But Michael and Lennart had the courage to think bigger, and their response built on ESA's request for interferometry.

### 15. GAIA and Gaia

The major advance was the proposal led by Lennart Lindegren and Michael Perryman to use a small-baseline interferometer, housed within the Ariane 5 fairing, to achieve better accuracy and bigger mirrors to get to fainter magnitudes. Although the idea of an interferometer was later dropped due to technical complexity, the advantage of the proposal was that interferometry had been foreseen within the Horizon 2000 cornerstone plans, and this really led to the mission gaining support; both because of its science goals, but also because of its technological appeal - even though other interferometry groups, especially infrared science - were still pushing their own ideas.

The reply to the ESA call for proposals of Cornerstone studies was submitted on 12 October 1993, a proposal to study for astrometry "a large Roemer option and an interferometric option", GAIA, Lindegren et al. (1993). They should be studied as two concepts for an ESA Cornerstone Mission for astrometry "without a priori excluding either one" as Lindegren wrote in the cover letter. But in those years, we soon considered interferometry to be very good for our purpose and the Roemer option was forgotten for several years.

In fact, I was also soon able to "think big" and proposed in August 1994 a larger Roemer mission, called Roemer Plus (Høg 1995), with larger apertures to obtain the high accuracy. It used direct imaging and no interferometry. But it attracted no interest because everybody, including myself, had come to believe that interferometry was the right way to go. But it was NOT AT ALL, as industry showed us in January 1998!

In 1995 I proposed a new optical design GAIA95 with Claus Fabricius and Valeri Makarov, it was a Fizeau interferometer with Gregorian telescope, providing dispersed fringes for simultaneous astrometry and spectrophotometry. This design was immediately adopted for a minisatellite project DIVA by the German space agency DLR. The project was started and led by Siegfried Röser, Uli Bastian and Elena Schilbach and was joined by US participation, but it was abandoned in 2002 after seven years when one of the German funding partners had dropped out. Luckily so we must say, because our German colleagues could then come back and join efforts to develop Gaia.

Ten years ago, I have told the two fascinating stories of "Roemer and Gaia" and of "Interferometry from Space: A Great Dream" in two reports, Høg (2011c, 2014d).

### 16. Scientific objectives

Besides of making technical contributions to Gaia, I also looked at some specific scientific objectives. One was figuring out what magnitude Gaia had to reach to measure specific Galactic populations.

I prepared a table of "Some Galactic kinematic tracers" which is placed as the first table in the Gaia Study Report, ESA (2000b). It was elaborated during the years since the earliest discussions I had in 1994 at the IAU Assembly in the Hague especially with Gerry Gilmore. It contains selected tracers with the limiting magnitude and astrometric information necessary to study them, and it "demonstrates that GAIA will provide adequate precision to meet the scientific goals". The selected tracers belonged to all Galactic populations: clusters, disk, halo and bulge, and the table was useful to justify the 18-20 mag completeness limit.

Another scientific idea was on astrometric microlensing and what could be done with Gaia. The possibility to detect both photometric and astrometric effects of microlensing was discussed in the ROEMER proposal in 1993. The effect of relativistic gravitation appears when a massive foreground object, called a MACHO, passes in front of a star further away when this star is observed close to the crossing, and given the billions of photometric and astrometric observations by the proposed mission, the effect might be detectable. Such observations could potentially be used to determine the mass of the foreground object, the MACHO.

In 1994 I had a collaboration with Russian cosmologists and showed (Høg, Novikov, Polnarev, 1995) that all six physical parameters of a MACHO (mass, distance, two position coordinates and two proper motion components) could in principle be derived from observations of an encounter event with a star for which the position and proper motion are known.

This was a nice theoretical result, but the practice is difficult. However, in fact twenty years later, the detection of events with Gaia and photometric follow-up observations from a network of telescopes on the ground has become a rich source of science. Over a thousand transient sources had already been detected with Gaia data up to October 2016.

### 17. Optimizing Gaia

I always wanted to optimize the instrument.

One problem was how to best sample the pixels of the CCD focal plane, and I contributed many technical notes on this over several years. The problem was that not all the billion CCD pixels could be sampled and sent to ground, because the read noise would be excessive, and the telemetry rate was limited. So, a clever selection of a subset of pixels was required onboard.

Much of the sky is empty, even at 20-21 mag so no such samples should be transmitted. But there are other complications from diffraction spikes, and double and multiple stars, and the sky is not so empty inside clusters and galaxies. All this had to be optimized so that the best science could be extracted from the data in powerful computers on the ground by clever astronomers and technical experts.

We also took care that the environment of all detected objects could be mapped in an area of about one square-arcsecond to detect companion stars or a nebulosity around the object which might itself be a compact galaxy or a quasar. This was studied in 1998-1999 by a student who visited me in Copenhagen from the University of Padua, Mattia Vaccari, as part of his MSc Thesis. Mattia is now a professor at the University of Cape Town. The thesis says that an all-sky Gaia survey of the central regions of galaxies brighter than I $\simeq$ 17 mag will find more than 3 million galaxies and observe in at least 4 colors with a spatial resolution better than 0.4 arcsec. In fact, the Gaia Data Release 3 claimed in 2022 that 2.9 million probable galaxies (with ~95% purity) have been found. The spatial resolution in the optical domain will be unprecedented, about 0.18 arcsec by the end of the mission.

Much of the work on sampling was done in collaboration with Frédéric Arenou (from Paris) and Jos de Bruijne (working under Michael in ESA), and the work was recorded in a series of technical reports. There was a total of 64 reports since 1997 with authors from Copenhagen about sampling, detection, and imaging for Gaia. The final sampling used in the satellite operations built on our work but was further optimized by industry and especially Jos de Bruijne at ESTEC. - Recently, I was very impressed and delighted seeing for the first time the optimization of these processes by de Bruijne et al. (2015), a depth of study I could not have imagined when I worked on this 20 years ago. They have optimized the rejection parameters in the detection process, improving - with respect to the functional baseline - the detection performance of single stars and of unresolved and resolved double stars, while, at the same time, improving the rejection performance of cosmic rays and of solar protons, etc. etc.

Another of my major activities in optimizing the Gaia payload was on the photometry. From the earliest ideas for Gaia, it was very clear that measuring a billion or more objects was made far more powerful if we could also record their colors or spectra. Photometry was important, in fact for two different reasons. The instrument gives small but significant chromatic errors in the astrometry which can be corrected if a color index of the star is known, and we need to know that for every transit for the sake of variable stars with changing colors. The other reason is astrophysical: the color gives its position in the Hertzsprung-Russell diagram, and (if the filters are carefully chosen) also the spectral type, metallicity, and reddening. Many other photometric surveys are going on, but we needed a reliable source giving prompt results: So, it had to be by Gaia itself.

There were several key people helping to think through the objectives, and the possible solutions. My colleague in Copenhagen, Jens Knude, was my constant support and he brought his lifelong dedication to photometry, especially with the Strömgren uvby-system with intermediate width bands suited for early type stars. People in Lithuania were also keen to support, in fact Professor Vytautas P. Straizys (1936-2021) wrote to me already in 1994. He was director of the Vilnius Observatory and had developed the "Stromvil" system suited for classification of all spectral types, also in the presence of interstellar reddening. So, the experts joined our efforts very early. For photometry and spectroscopy, the first idea was to add a third telescope with a collecting aperture of 0.75 x 0.70 m$^2$ in addition to the two large telescopes for astrometry. This was the design by Matra Marconi Space at the end of the industrial study in mid-1998.

Various possibilities were considered, ranging from multiple filters to low-resolution spectroscopy. When the cornerstone study started, a Gaia Science Team was formed in 2001, under Michael's leadership, with various working groups devoted to studying specific aspects. One of these was the photometry working group with Carme Jordi from Barcelona and me as co-leaders. We had many special meetings on photometry and wrote many reports in those years. In the first design, photometry in four broad bands was obtained in the astrometric telescopes and in seven medium-width spectral bands in the smaller telescope. The next design, which was described in detail in The Concept and Study Report, ESA (2000b), increased this to eleven intermediate bands. Our work ended with a paper (Jordi et al. 2006) of 25 pages in the journal Monthly Notices of the RAS with 35 authors. The number of bands had increased to a total of 19 optimized for astrophysics.

But that wasn't the end of this long and important story. NO! All this changed dramatically. In fact, it changed while we were finalizing the MNRAS paper, but only a few of us knew that. The change had to be kept confidential because it was taking place during the competitive bid for the industrial contract for Gaia. One of the competitors, Matra Marconi Space (who eventually became the winning contractor), was proposing to simplify the design for reasons of mass and cost. They dropped the dedicated photometric telescope and reduced the number of focal planes from four to only one. (I guess that they have seen this possibility earlier but kept it for themselves until a strategically better moment of time.)

They asked Michael whether it would be acceptable to replace photometric filters with two low-dispersion prisms. They reluctantly accepted that he also engaged me, and I also wanted to have Carme Jordi, Anthony Brown from Leiden, and Lennart Lindegren involved. It was a complicated problem, but eventually a very careful optimization of the prism properties was done utilizing the MNRAS paper, resulting in the very satisfactory Gaia photometry that we have today. Radial velocities are obtained with a separate high- dispersion spectrograph, see figure 17.

I believe, however, that our decision would have been different if we had known much earlier that the dedicated photometric telescope had to be dropped. Given more time to think, we would probably have chosen multi-color (4 or 5) filter photometry for Gaia because the resulting high spatial resolution would give better photometry at multiple stars and in regions with high star density and the photometric accuracy would be better because of the lower background than with spectra. But that was not an option in 2006 at this critical point in the development. A 4-color broad-band photometry was in fact included in the design on figure 3.8 of the astrometric field in ESA (2000b) and this was improved to five bands in the final report on Gaia photometry (Jordi et al. 2006). – For a future mission as discussed in section 20, a combination of filter photometry and spectra is perhaps optimal in view of the very good experience with Gaia photometry, but this can hardly be accepted because the extra space needed in the focal plane if both filters and spectra are included would necessarily be taken from astrometry. Furthermore, 1) an advantage of filters is that the photometric observations can also be used for astrometry and 2) it is presently not clear which advantages for astrophysics low-dispersion spectra in the NIR might have over filters.

The Gaia mission was developed, the satellite was launched in December 2013 and this wonderful instrument is shown in the figures 17, 18 and 19.

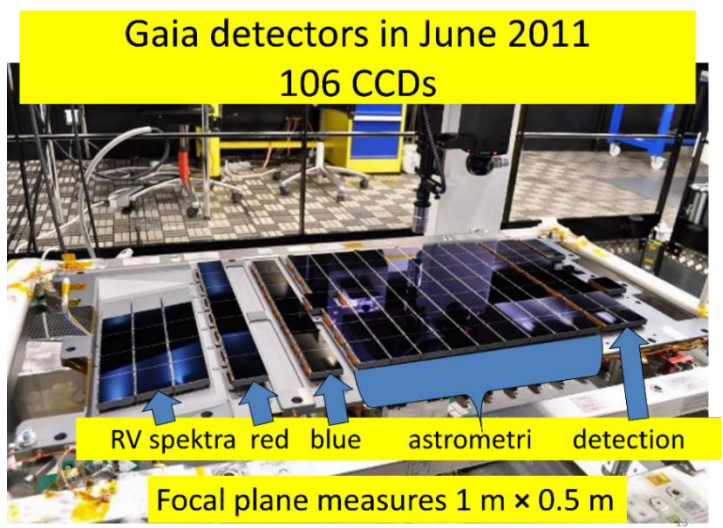

*Figure 17. The 106 CCD detectors in the focal plane of Gaia. Stars will be moving from right to left. They are detected and confirmed in the first two strips of CCDs and then measured for astrometry in the next nine strips. Low dispersion prisms, not shown, generate short spectra measured at 'blue' and 'red'. A high dispersion spectrograph generates spectra measured for radial velocity, RV. This was the largest area of CCDs ever on a science mission in space when Gaia was launched in 2013 and it still is; Euclid VIS has "only" 609 million pixels. The sophisticated, custom-built charge coupled devices (CCDs), are light detectors of essentially the same kind as found in a digital camera. Containing 106 CCDs, the focal plane assembly comprises a total of nearly one billion pixels (a 'gigapixel'), compared to the few million of a typical digital camera. - Figure from the website of ESA.*

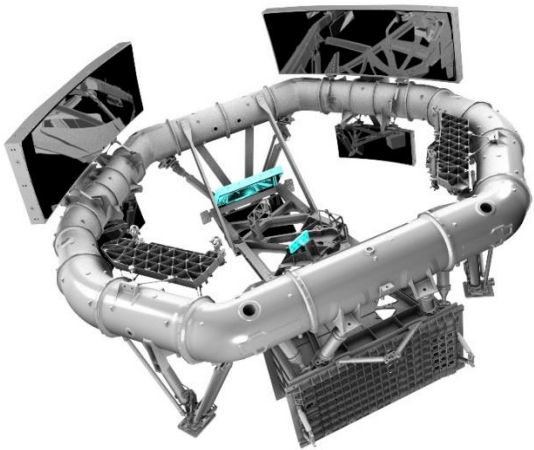

*Figure 18. Gaia mechanics and optics. A very stable torus of expansion free silicon carbide carries the optical components and the focal plane assembly. The two primary mirrors at top also of silicon carbide measure 1.45 × 0.50 m² and are covered with silver for high reflectivity. The two telescopes sharing a common focal plane, are each looking out through an aperture in the payload housing and are separated by a highly stable basic angle. Light from a celestial object enters the arrangement through one of the two apertures, striking the large primary mirror opposite. The light is then reflected by a series of mirrors along a total focal length of 35m, with the two light paths meeting at a beam combiner before finally reaching the shared focal plane. - Figure from the website of ESA.*

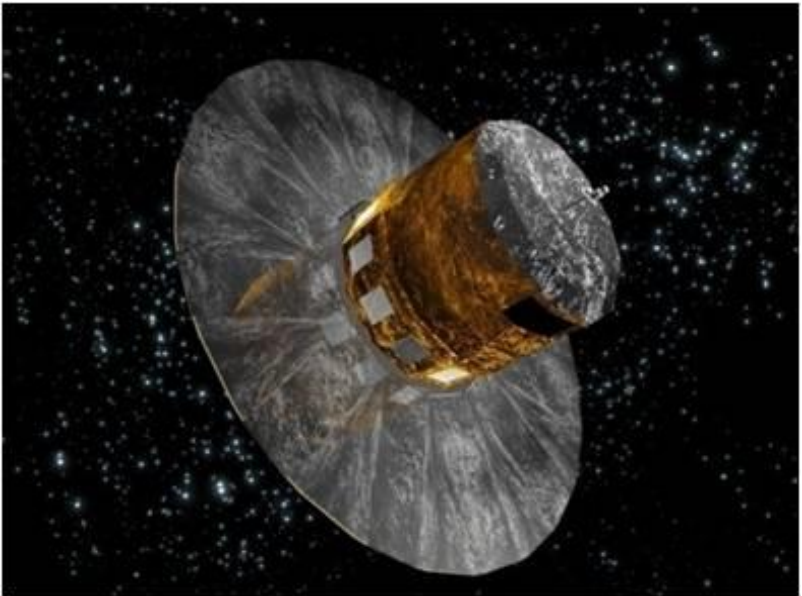

*Figure 19. The Gaia astrometric satellite launched by ESA in December 2013. The large circular sunshield (10.2 m diameter) at left gives permanent shadow so that no direct light from Sun, Moon or Earth ever reaches the thermal tent (which has a cylindrical shape) with the optics and electronics providing a stable and low temperature environment Preventing thermal instabilities is crucial for the mission's scientific objectives as they would affect the final accuracy of the micro-arcsecond astrometry measurements. The rectangular opening for one of the two telescopes is seen black at the right. - Figure from the website of ESA.*

## 18. Leaving Gaia in 2007

In 2007 I left the Gaia Science Team because ESA set up a new team. It had been a long journey for me since 1975 when I first became involved with ESA and space astrometry, full of work and pleasure. I could see that my work was done, and I could hand over to others. No Danish involvement in Gaia data reduction was conceivable. I was 75 enjoying my pension, and the institute was entirely focused on astrophysics, which I find very reasonable. Claus Fabricius and I are therefore the last Danish astrometrists in the row beginning with Tycho Brahe.

The Copenhagen University Observatory no longer exists. It started on top of the famous Round Tower in Copenhagen built in 1642 where Christen Sørensen Longomontanus, Tycho Brahe's assistant on Hven, had begun his work. And the Brorfelde Observatory, 50 km from Copenhagen, where I made the very first observations with the new meridian circle in 1953, was closed for science when the whole staff moved to Copenhagen in 1996. In 2005 the 363-year-old Copenhagen University Observatory was absorbed in the Niels Bohr Institute - without any celebration or funeral.

I have maintained a big interest in Gaia, and indeed in the role and future of space astrometry, I shall come to that. I am doing nearly all my work from home, but I kept an office in the institute until 2019 and often visited the cosmology group for talks. I wrote a popular article in Danish together with two cosmologists, Peter Laursen and Johan Samsing, about the beginning of the Universe, and more recently an article focused on explaining the cosmic horizon. I have asked the opinion of many astronomers in *interviews about an infinite universe*, Høg (2014e). I am happy that Johan Fynbo and his student Kasper Heintz adopted and implemented my idea of detecting quasars by their zero proper motion using Gaia results, Heintz et al. (2018).

I have given a lot of talks around the world about my life with astrometry, and I wrote many reports about the history of astrometry. I talked in Stuttgart in 2019, when I received the instrumentation prize from the German Astronomical Society together with Lennart and Michael, shortly after I spoke in Paris and in Berlin. I've given other talks online, in 2020 first to an audience in Potsdam, then to Perth in Australia invited by my colleague Andreas Wicenec from the Tycho-1 and Tycho-2 projects when he worked in Tübingen, and to South Africa invited by Mattia Vaccari who came from Padua and was my student in Copenhagen in the 1990s working on the mapping of galaxies with Gaia data.

### 19. Launch of Gaia and after

At ESA's invitation as a VIP, I had flown out to see the launch of Hipparcos by Ariane 4 from French Guiana in 1989. I had contributed to the developments underpinning its successor, Gaia. I was greatly disappointed that ESA did not recognize my contributions to this next step when deciding on the attendees at its launch from French Guiana in December 2013.

To my great delight, however, the day of Gaia launch came to please me very much in Copenhagen. My colleagues, especially Uffe Gråe Jørgensen, had arranged a meeting for the occasion at the Niels Bohr Institute (where I had my office) with invited guests, talks and transmission of the launch. I had invited several guests, my wife Aase was there, and my admired tutor from the 1950s, Professor Peter Naur, whom I talked about before. After a short talk at the meeting, I went with Peter Naur in a taxi to the TV studio where we both appeared for interview during the launch. You cannot imagine how happy I was to see him enjoying this event with me, his student from sixty years ago! I should add that we had frequent contact on the telephone, until he died a few years later.

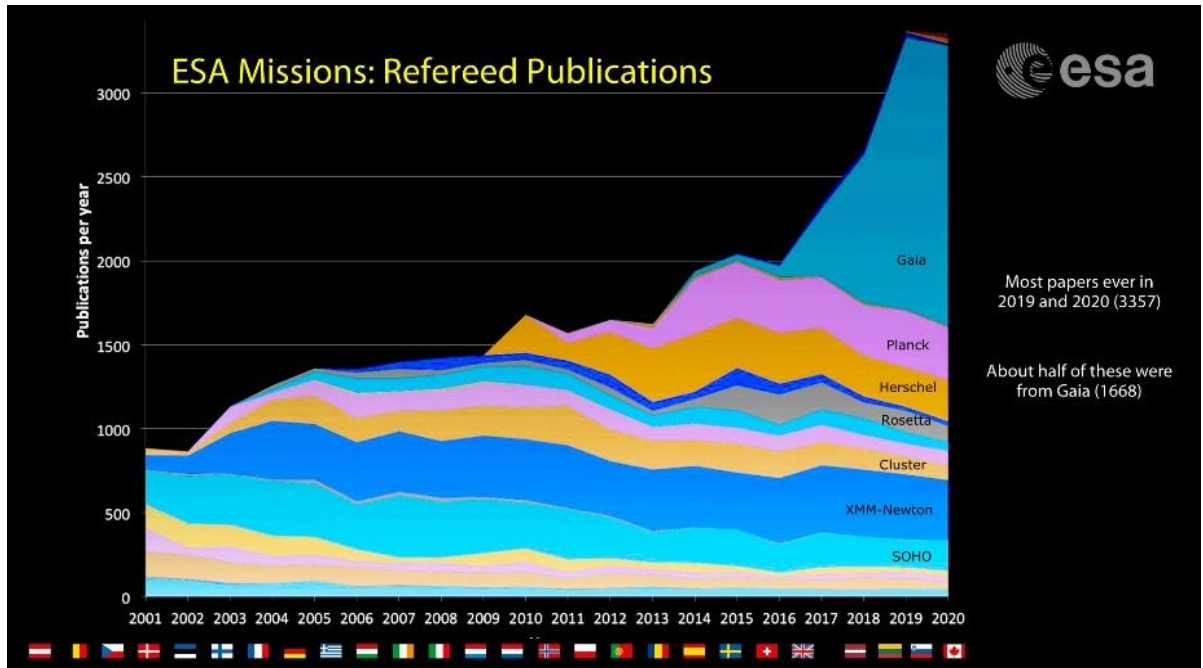

*Figure 20. Publications from ESA missions during 2001-2020. Gaia has in 2019 and 2020 given as many scientific publications per year as all the other ESA missions together. It appears that a revolution in all branches of astronomy and astrophysics was happening already from the first 22 months of observations. - Figure from the website of ESA.*

Gaia was launched at the end of 2013, and operations started in mid-2014, after a 6-month commissioning phase. The original target mission duration was for 5 years, but with consumables on-board for 10 years, and it seems to hold. Thanks to the superb mission operations team in ESOC, and the science operations team in ESAC, Gaia has now acquired more than 9 years of high-quality data with five Data Releases issued so far - in 2016, 2018, 2020, June 2022 with DR3, and the most recent FOcused Product Release on 10 Oct. 2023. The impressive science result is illustrated in figure 20.

It is overwhelming to see what has come out, and that the data reduction and everything has been able to follow up. Very good management, very good people. I have listened to Michael's interview with Anthony Brown, leader of the data reduction consortium, it touched me deeply, almost to tears. It is a wonderful and dedicated team, more than 400 people working hard to get these wonderful results out. How lucky I am to live long enough to see all this myself. – I am still able to do something useful for astronomy, so it seems from section 20.

## 20. The Gaia Successor in view

In May 2013 I heard from Claus Fabricius (in Barcelona) about ESA's call for new Large Missions, just 11 days before deadline. The call was a surprise, but it was immediately clear to me that it should be a Gaia successor in 20 years. When writing the proposal I corresponded with colleagues, some urged me to submit it, a few others said I should not submit because it was not ripe and because such a proposal should be submitted by a committee, not by a single person. But I had only a few days to write and to decide. I submitted the proposal in time, which was six months before Gaia was launched, Høg (2014a).

I was sure the mission should use filter photometry with the same high angular resolution of 0.12 arcsec as the astrometry for the benefit of narrow multiple stars and dense fields of the sky where multicolor photometry of the individual stars can only be obtained with filters. This will be most important for *GaiaNIR*, the name now used for the successor, where the star density will be even higher than in Gaia.

Prism photometry as in Gaia with a resolution about one arcsec should not stand alone. The prism photometry of Gaia has been a great success but a quantitative comparison with filter photometry in a Gaia successor should be made, see Høg (2023b). My instinct tells me that the mission will end up without low-dispersion spectra, but the question must of course be carefully studied. A medium-dispersion spectrograph shall provide radial velocities which are very desirable to obtain in larger number.

The idea behind this second Gaia was that astrometric observations of the Gaia stars after 20 years would give ten times more accurate proper motions from positions at the two epochs than from a single mission. The benefit for kinematic studies would be immense for more distant stars, and for stars close-by due to better accuracy, and for maintaining the celestial reference system. Exoplanets with long Jupiter- and Saturn-like periods could be detected and the whole population of such systems in our neighborhood could be discovered. During the following years I wrote a long report (Høg 2014b) with all the scientific cases while corresponding with many colleagues and I presented the ideas at many meetings. I had response from at least a hundred competent people in oral discussion or correspondence.

With Jens Knude and other colleagues in Copenhagen I began in 2014 to look at the possibility to extend the wavelength range into the near infrared, beyond the 1000 nm possible with the classical CCDs. The benefit would be the observation of redder stars, i.e., cooler stars or stars dimmed and reddened by dust clouds.

In 2015 a colleague at the Lund Observatory, David Hobbs, joined with much energy and he is now the natural leader of the project. We focused on GaiaNIR, a successor in 20 years with a Near-Infrared capability, Hobbs et al. (2021, 2023). A technical study was conducted by ESA, reported in ESA (2017). A design of the optical system is shown in figure 21. *For the Concurrent Design Facility (CDF) study 2 entrance pupils of 1.6x0.25m were used. These values are not fixed. The maximum AL size is 1.7m without involving expensive manufacturing procedures. It is still being considered if the AC size can be doubled to 0.5m.*

The NIR detector is critical since we need TDI, time-delayed integration, which was easy to do in the visual with CCDs in Gaia but is difficult in NIR. Most of the stars observed by Gaia shall be observed again by the successor since a good sensitivity at less than 1000 nm shall be maintained. The positions at two epochs will be obtained for them and thus the highly accurate proper motions.

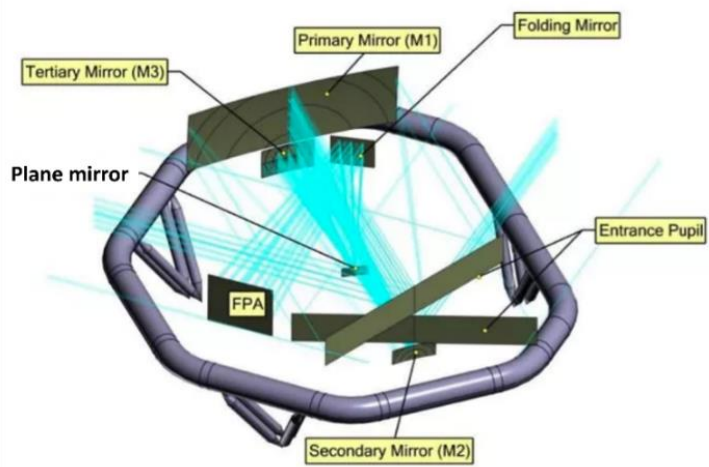

Figure 21. GaiaNIR, the Gaia successor planned to be launched in 2045: Optical surfaces and the light path from the ESA CDF study. This design differs from Gaia's in that: the mirror surfaces are simple conics to simplify manufacturing, alignment and test, and the entrance pupil is at a flat folding mirror in front of the primary instead of on the primary mirror itself. – Figure based on the GaiaNIR CDF study report ESA (2017).

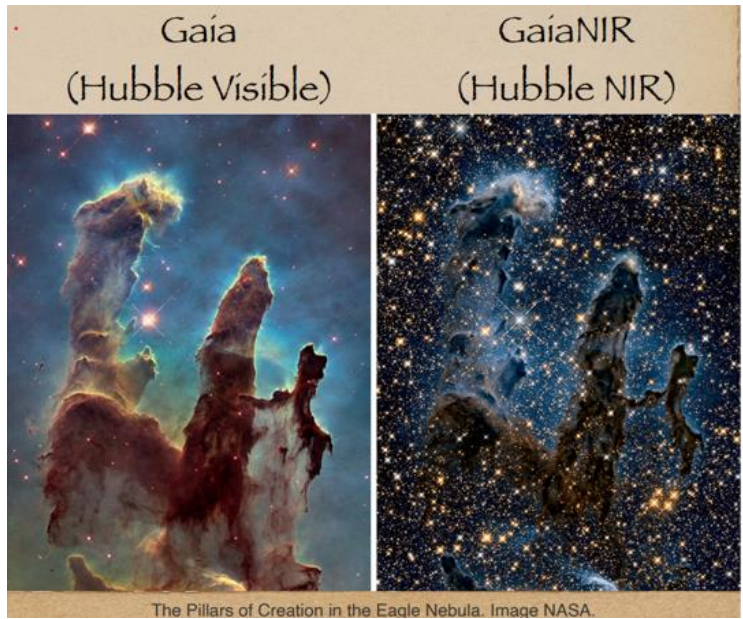

*Figure 22. A Gaia successor was proposed in 2013 (Høg 2014a) to measure the Gaia stars again, twenty or more years later giving much more accurate proper motions when combining the positions from the two epochs. This project has developed to include sensitivity in the NIR, near infrared, so that stars in obscured region and redder stars can be reached together with most of the stars observed by Gaia. The two Hubble pictures of the Pillars of Creation in the Eagle Nebula illustrate the enormous gain. - Slide by D. Hobbs, using images from the website of NASA.*

An industrial study of NIR detectors for our project is making progress, perhaps with sensitivity up to 2500 nm. The new mission will be able to peer into the obscured regions of the Galaxy (figure 22) and measure up to 10 or 12 billion new objects, in addition to two billion Gaia stars, and reveal many new sciences in the process. ESA has now ranked the development of this mission so high that it has good prospect of being realized and launched around 2045, see Høg (2021).

Many years ago, in 2005 I wrote: "The Gaia astrometric survey of a thousand million stars cannot be surpassed in completeness and accuracy within the next forty or fifty years." I knew how long it would take to get a large new mission approved and developed, and we can expect the first results from GaiaNIR in 2050 if all goes well - which is within the time frame I mentioned. The mission will supersede Gaia in four respects: The infrared capability, the 10 billion stars, the extremely accurate proper motions from positions at the two widely separated epochs of Gaia and GaiaNIR, and the two missions of 10 years will give well-defined samples of multiple systems with astrometric orbits of up to 30-year periods and longer, including stars and exoplanets.

With good luck I have been able to help bringing this bright future within reach. It is like a dream, but it seems to be true! It has happened so soon because the proposal with a realistic design was submitted in May 2013 even before Gaia had been launched and work was started on the project. Two years later, David Hobbs became active for the Gaia successor. David has worked on Gaia since 2007 and he brought great scientific and technical insight, and he has made full use of all the possibilities offered by ESA.

Another response to ESA's call in 2013 for mission proposals was submitted by Anthony Brown as leader of a group of astronomers, Brown (2013) "Space-Time Structure Explorer, Sub-microarcsecond astrometry for the 2030s" where the great scientific gains were detailed. No mission concept was presented, but the challenges on the road for an all-sky mission were analyzed and NIR sensitivity was considered.

In June 2022 Anthony Brown sent his congratulations to my 90th birthday which was celebrated at the Lund Observatory: "… I hope you like our little present in the form of Gaia Data Release 3 [which had just taken place]. Now that we have clearly started looking forward to GaiaNIR, I want to thank you for your insistence that next to dreams of nano-arcsecond astrometry we should also be thinking about realistic alternatives. Your push for a second Gaia in 2013 got the ball rolling on GaiaNIR and was the first step to come to where we are today.

I listened with great pleasure to your interviews with Michael Perryman. One remark you made really struck me. Commenting on your career path, you said that through Hipparcos and Gaia you had done more for astrophysics than you could ever have done as an astrophysicist. A remark I certainly take inspiration from!

I wish you all the best and a wonderful day in Lund. - Anthony, Dr. Anthony G.A. Brown, Leiden Observatory ".

From Michael Grewing came: "…my congratulations and my very best wishes for this day and for the years to come! I have now known you for almost 60 years, and I am still deeply impressed by your enthusiasm, your skills, and your leadership, by convincing, in a very efficient manner, others what is the next step to go, even if this "next step" is sometimes very far in the future.

What could have been a better proof of your manifold achievements than yesterdays' 3rd release of data from the GAIA mission! This is a unique milestone in the history of astrometry, which might never have come true, at least not in such a timely manner after the HIPPARCOS mission, without your voice being heard, clearly and persistently over the years! I hope this will continue to be so!

Thank you for pointing out to me yesterdays' event and the gathering today to celebrate your anniversary! I will try to join the zoom meeting if I can. My thoughts will in any case be with you all this day."

Andreas Wicenec recently wrote:"...I always regarded you as my de facto supervisor [beginning the PhD in 1987 in Tübingen] for most of the work I had been doing, in particular within the Tycho consortium. I learned so many things from your way of working and during my discussions with you about supposedly simple plots, captions, and paragraphs. I still recommend similar things to my students, always having yours in mind. You have been, and still are, absolutely fundamental to astrometry in general and to all the people around you as well."

Nicholas Rowell wrote after reading a draft of this paper: "...It has been fascinating to learn about the long road to Gaia (and beyond) and the immense efforts made over many years by yourself and some of my Gaia colleagues in achieving that. It should be required reading for all 'young' Gaia people like me who have joined the party relatively recently."

More recently, the referee of the present paper began his report thus: "My total impression is that it is a very readable and attractive account of the colossal developments in a long period for astrometry. A very valuable asset for history.

The state-of-affairs has changed from very marginal data on stellar distances and stellar motions to the present still developing catalogue of high-precision data on almost $2 \times 10^9$ objects! The statistical weight of the present knowledge is an astounding factor of almost $10^{12}$ larger than what had been achieved by ground-based efforts in astrometry since say early 19[th] century, and the developments leading to this are well described by a key active participant in essentially all this work."

The referee has now concluded: "It is a very fine comprehensive account of how such a colossal progress came about from a status of just usable hard to get data as it was. The new situation is almost to be taken for granted by the young astrophysicists of today.

The process of elaborating the new science possible with this progress is still ongoing and will take yet a few more decades for the full harvest."

Lund Observatory invited scientists to join a meeting in July 2023: "Science and technology roadmap for μas studies of the Milky Way" where about 80 registered and some attended offline. I gave a short presentation (Høg 2023a) and enjoyed listening and talking with the colleagues. That gave me some ideas about how TDI mode could be approximated by on-board processing of the detector frames read at pixel passage rate, and how this could be optimized even for photometry, sent in a note to some and elaborated in correspondence: Høg (2023b). Lennart wrote about my new idea for the first onboard treatment of the detector data: "I think it is an ingenious idea if it can be implemented. Reducing the across-scan smearing to an almost insignificant level will increase sensitivity, reduce image overlap, and simplify the PSF calibration." The note also discusses the multicolor photometry with GaiaNIR.

## 21. Closing off

I was very happy about the recognition from the German Astronomical Society mentioned above, and especially that I received it together with Michael and Lennart. We three have enjoyed a wonderful collaboration for so many years, with Lennart it began in 1973 almost 50 years ago, and with Michael in 1981.

Recently I said to Michael: "When the European Astronomical Society announced its new prize in 2008, the Tycho Brahe Medal, I thought secretly, that I might get it! But when you then finally got it in 2011, I thought: *that is perfectly right*, Michael deserves it for his brilliant leadership of the Hipparcos and Gaia Teams since we four consortia leaders of Hipparcos had already received the Director of Science Medal from ESA in 1999."

Let me recall what we wrote in the proposal for the ROEMER mission 30 years ago: "*The mission is not primarily aimed at solving some particular problem, but rather to explore the rich complexity of Nature in order to find new patterns, anomalies and even contradictions to existing concepts*". This is also the spirit of the proposed GaiaNIR mission, and this vision has become true in a spectacular way for the Gaia mission when we see in figure 20 that the number of scientific papers per year has been larger since 2019 than for all the other ESA missions taken together, and only based on the first 22 months (DR1 and DR2) of Gaia observations. Imagine how it will look with observations from 10 years=120 months!

Accuracy of measurement is needed in science, and this has been greatly improved in astrometry for twenty centuries, the position errors today with Gaia are 100 million times smaller than in Antiquity as illustrated in figure 23. I have reported many times on this progress, most recently in Høg (2020) where the present figure is given.

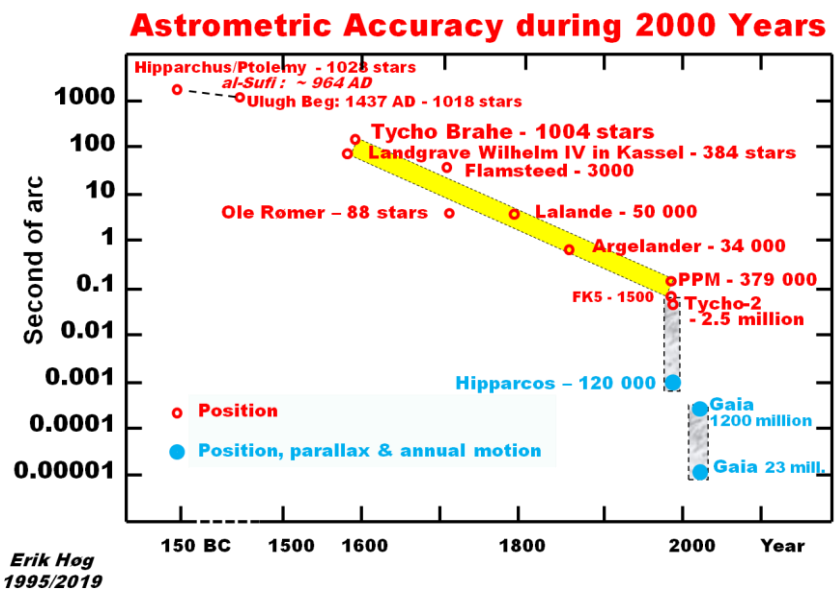

*Figure 23. Astrometric accuracy for 2000 years. The accuracy was greatly improved shortly before 1600 AD by the Landgrave in Kassel and by Tycho Brahe in Denmark. The following 400 years brought an even larger more gradual improvement before space techniques with the Hipparcos satellite started a new era of astrometry. In this 30-year era the errors have been reduced 10,000 times, as much as in the preceding 2000 years since Antiquity. – Ground-based parallaxes are omitted here for simplicity, but they are included on figure 1 in Høg (2020). – Figure copied from Høg (2020).*

Allow me now to leave astronomy for a moment and speak about my concern for our civilization which is shared by many people. NOT the planet Earth is in danger as we often hear in discussion of climate and global warming, because plants, animals and humans will of course survive and adapt to the changing conditions. It is *"only"* human civilization that will suffer increasingly. The living conditions have been improving for a large fraction of people on Earth during the past centuries, but this will not continue for very long for many reasons: increasing pollution of atmosphere, soil and oceans, people smuggling, international crime, and because our political structures are probably unable to fundamentally reverse this trend.

But 500 years ago, Martin Luther spoke a good motto for everyone:

*"Even if I know that the world will perish tomorrow, I will plant a tree today."*

## 22. Acknowledgements

I have emphasized that this article focuses on my own involvement and experiences over my lifetime. Making Hipparcos and Gaia the successes that they are obviously involved many other individuals, many institutes and organizations across Europe, numerous technical and computational complexities, and many other political, organizational, and technological challenges. I am happy I can quote much directly from the interviews by Michael Perryman in Høg (2022), and I am grateful for further information and comments from Uli Bastian, Anthony Brown, Jos de Bruijne, Claus Fabricius, Michael Grewing, Jean-Louis Halbwachs, David Hobbs, Povl Høyer, Ningsheng Hu, Carme Jordi, Zhigang Li, Valeri V. Makarov, Lennart Lindegren, Michael Perryman, Nicholas Rowell, Natalia Shakht, Catherine Turon, Mattia Vaccari, Andreas Wicenec, Roland Wielen, Yoshiyuki Yamada, Vladimir Yershov, Dongshan Yin, and from the referee. I thank Andreas, Anthony, Lennart, Michael Grewing, Michael Perryman, Nick Rowell, Roland Wielen, Yoshiyuki, and also the referee for their kind permission to quote. The referee for this publication was Prof. Rudolf S. Le Poole (Leiden University), who kindly agreed to reveal his identity, and provided great suggestions for the improvement of the manuscript.

I am indebted to the Royal Astronomical Society for awarding me an Honorary Fellowship for "a distinguished career over nearly seven decades…" which has led to the invitation to write the present review. This review contains in essence what Professor Rajesh Kochhar (1946-2022), president of the IAU Commission for History, fourteen years ago repeatedly urged me to write as a book: *my scientific biography*.

## 23. References[2]

---